\documentclass[12pt]{article}
\usepackage{makeidx}
\usepackage{amsmath,amssymb}
\usepackage[dvipdfm]{graphicx}
\usepackage{color}
\usepackage{bm}
\usepackage{authblk}
\usepackage{here}
\usepackage[font=footnotesize]{caption}
\usepackage{cite}
\usepackage{caption}

\definecolor{DarkBlue}{rgb}{0,0,0.7} 

\definecolor{DarkRed}{rgb}{0.65,0,0} 
\newcommand{\dred}[1]{{#1}}
\definecolor{DarkGreen}{rgb}{0,0.6,0} 
\newcommand{\magenta}[1]{{#1}}

\title{Expanding universe with nonlinear gravitational waves}
\author[1]{Taishi Ikeda\footnote{Electric address:ikeda@gravity.phys.nagoya-u.ac.jp}}
\affil[1]{\it{Departure of Physics,~~Graduate School of Science,~~Nagoya University,~~Nagoya~~464-6602,~~Japan}}
\author[1]{Chul-Moon Yoo\footnote{Electric address:yoo@gravity.phys.nagoya-u.ac.jp}}
\author[1]{Yasusada Nambu\footnote{Electric address:nambu@gravity.phys.nagoya-u.ac.jp}}
\date{}
\begin{document}
\maketitle
\begin{abstract}
We test the validity of Isaacson's formula 
which states that high frequency and low amplitude gravitational waves 
behave as a radiation fluid on average. 
For this purpose, we numerically construct a solution of 
the vacuum Einstein equations 
which contains nonlinear standing gravitational waves. 
The solution is constructed in a 
cubic box with periodic boundary conditions. 
The time evolution is solved in a gauge in which the trace of the extrinsic curvature $K$
of the time slice becomes spatially uniform. 
Then, the Hubble expansion rate $H$ is defined by $H=-K/3$ and 
compared with the 
effective scale factor $L$ defined by the proper volume, area and length of the cubic box. 
We find that, even when the wave length of the gravitational waves is 
comparable to the Hubble 
scale, 
the deviation from Isaacson's formula $H\propto L^{-2}$ is 
at most 3\% without taking a temporal average and is below 0.1\% 
with a temporal average. 
\end{abstract}
\section{Introduction}\label{introduction}
The global aspect of our universe is well described by 
a homogeneous and isotropic FLRW(Friedmann-Lema$\hat{\rm{i}}$tre-Robertson-Walker) metric. 
The homogeneity over the scale of 100Mpc is well established 
by many observations. 
At the same time, 
local inhomogeneity of our universe provides us a lot of information. 
One of the classical issue involved with the inhomogeneity 
is the so called backreaction problem(see e.g. \cite{Ellis:2011hk,Green:2014aga,Nambu:2000ex,Kai:2006ws}). 
Evaluation of the backreaction 
due to 
the local inhomogeneity 
on the global expansion law has been attracted much attention 
because large backreaction may cause significant systematic error 
for evaluation of the energy components of our universe. 
\par
One typical example of treating 
the backreaction is Isaacson's formula\cite{Isaacson:1967zz,Isaacson:1968zza}. 
For gravitational waves with high frequency and low amplitude, Isaacson's formula provides the coarse-grained effective energy momentum tensor which satisfies the traceless condition. 
Since the energy momentum tensor which is compatible with the 
homogeneity and isotropy is given by the perfect fluid form, 
Isaacson's formula states that 
the high frequency and low amplitude gravitational waves 
behave like a radiation fluid on average (see Appendix \ref{Isaacson_in_Freidmann}). 
Therefore, for a spatially flat FLRW universe with short wavelength gravitational waves, the relation between the Hubble parameter $H$ 
and the scale factor $a$ is expected to be $H\propto a^{-2}$. 
\par
The aim of this paper is to 
test the validity of Isaacson's formula beyond the 
high frequency and low amplitude approximation.  
For the purpose, we consider the inhomogeneous universe which only contains the gravitational waves and calculate the time evolution of this universe. 
Because this universe is assumed to have no symmetry, it is inevitable to rely on a numerical computation to obtain its dynamics and we apply numerical relativity to tackle this problem. 
Over the past 20 years, numerical relativity has been extensively developed mainly for isolated systems. 
Recently, 
several authors applied the numerical relativity 
to the expanding or contracting universe models with aligned black holes 
and discussed the backreaction problem\cite{Bentivegna:2012ei,Yoo:2012jz,Bentivegna:2013jta,Yoo:2014boa}. 
\par
In this paper, we 
investigate the validity 
of Isaacson's formula 
by solving the full Einstein equations with the use of numerical relativity. 
First, we construct 
the initial data 
which contains nonlinear gravitational waves 
by 
solving constraint equations of vacuum Einstein equations. 
This initial data 
is prepared in a cubic box with periodic boundary conditions 
between each pair of opposite faces of the box and 
corresponds to 
standing wave modes in the linear approximation. 
Concerning the time evolution, the backreaction effect of gravitational waves causes expansion or contraction of this universe. 
We use the gauge condition in which the trace of the extrinsic curvature $K$ is spatially uniform. 
Then, the Hubble parameter is defined by $H=-K/3$. 
We introduce the effective scale factors $L$ from the proper volume, proper area, and proper length of this box. 
Isaacson's formula states that time evolution of this system is same as the universe with the radiation fluid in the short wavelength region 
and the Hubble parameter and the effective scale factors obey the relation 
\begin{equation}\label{Isaacson formula}
H=\frac{b}{L^{2}},
\end{equation}
where $b$ is a constant. 
We evaluate the relation between the Hubble parameter $H$ and the effective scale factor $L$ in 
the numerically generated inhomogeneous universe and test the validity of this formula. 
\par
This paper is organized as follows. 
In Sec.~2, the initial data are constructed by solving constraint equations 
of the vacuum Einstein equations. 
We describe the time evolution of our model in Sec.~3 
and the relation between the Hubble parameter $H$ and the effective scale factor $L$ 
is investigated. 
Furthermore, 
we report dependence of the initial amplitude on 
the dynamics in superhorizon scale and the effect of temporal average of the volume. 
Sec.~4 is devoted to a summary and discussion. 
We use 
units 
in which 
$c=G=1$ throughout this paper. 
\section{Set up of initial data}\label{Set up of initial data}
\subsection{Construction of initial data}
As is mentioned in the introduction, 
we numerically construct a solution of the vacuum Einstein equations, 
which contains nonlinear standing gravitational waves in the cubic box 
with periodic boundary conditions. 
Naively thinking, since the gravitational waves have a positive energy, 
it causes the gravitational attractive force, so this system is unstable. 
Actually, as we will see later, this universe inevitably expands or contracts. 
In this section, we describe how to construct the initial data. 
\par
The initial data set is generated by solving 
the following Hamiltonian and momentum constraints: 
\begin{equation}
\mathcal{R}+K^{2}-K_{ij}K^{ij}=0,
\end{equation}
\begin{equation}
D_{j}K^{j}_{\ i}-D_{i}K=0,
\end{equation}
where $\mathcal{R}$ is the scalar curvature of 3-metric $\gamma_{ij}$, 
$K_{ij}$ is the extrinsic curvature, $K=\gamma_{ij}K^{ij}$, 
and $D_{i}$ is the covariant derivative associated with $\gamma_{ij}$. 
For convenience, $\gamma_{ij}$ and $K_{ij}$ are decomposed into 
the conformal factor $\Psi$, conformal 3-metric $\tilde{\gamma}_{ij}$, 
and conformal traceless part of the extrinsic curvature $\tilde{A}_{ij}$ 
as follows:  
\begin{equation}
\begin{array}{rclrcl}
\gamma_{ij}&=&\Psi^4 \tilde \gamma_{ij},&{\rm det} (\tilde{\gamma}_{ij})&=&1,\\
K_{ij}&=&\Psi^4\left(\tilde A_{ij}+\frac{1}{3}\tilde \gamma_{ij}K\right),&\tilde \gamma_{ij}\tilde A^{ij}&=&0. 
\end{array}
\end{equation}
Using these variables, the Hamiltonian constraint and the momentum constraints are written as 
\begin{equation}
\tilde{D}_{i}\tilde{D}^{i}\Psi-\frac{1}{8}\tilde{R}\Psi+\left( \frac{1}{8}\tilde{A}_{ij}\tilde{A}^{ij}-\frac{1}{12}K^{2}\right) \Psi^{5}=0, 
\end{equation}
\begin{equation}
\tilde{D}_{j}\tilde{A}^{ij}+6\tilde{A}^{ij}\tilde{D}_{j}\ln \Psi-\frac{2}{3}\tilde{D}^{i}K=0,
\end{equation}
where $\tilde{D}_{i}$ and $\tilde{R}$ are the covariant derivative and 
the scalar curvature associated 
with $\tilde{\gamma}_{ij}$, respectively. 
In order to construct the initial data, 
we solve these constraints with appropriate ansatzes. 
\par
We assume that $K$ is spatially constant 
and $\tilde{A}_{ij}=0$. 
These assumptions make momentum constraints trivial. 
Furthermore, the Hamiltonian constraint is reduced to 
\begin{equation}
\tilde{D}_{i}\tilde{D}^{i}\Psi-\frac{1}{8}\tilde{R}\Psi-\frac{1}{12}K^{2} \Psi^{5}=0. 
\label{Hamicon}
\end{equation}
We also assume that the conformal metric has the following form: 
\begin{equation}\label{3d-metric}
\tilde{\gamma}_{ij}=\mbox{diag}\left(\frac{1+h^{(3)}}{1+h^{(2)}},
\frac{1+h^{(1)}}{1+h^{(3)}},\frac{1+h^{(2)}}{1+h^{(1)}}\right), 
\end{equation}
where $h^{(i)}=\mathcal{A}^{(i)}\cos (\omega_{0} x^{(i)}+\phi^{(i)})$, 
($i=x,y,z)$\cite{Nakao}.
A constant $\omega_{0}$ specifies the coordinate wave number of the gravitational wave\dred{s} and 
$\phi^{(i)}$ is the phase of the gravitational waves.
As we will see later, in the linear approximation, these ansatzes lead to a solution for standing gravitational waves 
with the amplitude $\mathcal{A}^{(i)}$.
In this paper, for simplicity, we assume
\begin{equation}\label{set_amp}
\mathcal{A}^{(x)}=\mathcal{A}^{(y)}=\mathcal{A}^{(z)}=\mathcal{A}.
\end{equation}
By this assumption, three spatial axes are equivalent to each other. 
\par
We consider periodic boundary conditions for 
each pair of opposite faces of the numerical box. 
The coordinate length of the edge of the box is set to be 
$\lambda_{0} =2\pi/\omega_{0}$. 
\footnote{In practice, 1/8 region of the box with reflection boundary condition 
is enough because of the discrete symmetry. }
Under this boundary condition, 
without loss of generality, we can set 
\begin{equation}
\phi^{(x)}=\phi^{(y)}=\phi^{(z)}=0.
\end{equation}
\par
By integrating the Hamiltonian constraint over the box, we obtain 
\begin{equation}\label{Kval}
K=\pm \sqrt{-\frac{3\int_{\rm box} d^{3}x \sqrt{\tilde{\gamma}} \tilde{R}\Psi}{2\int_{\rm box} d^{3}x \sqrt{\tilde{\gamma}} \Psi^{5}}},
\end{equation}
where the first term in Eq.~\eqref{Hamicon} 
has been reduced to the boundary integral and 
eliminated due to the periodic boundary conditions. 
Since the trace part of the extrinsic curvature gives 
the volume contraction rate, 
our solution describes expanding or 
contracting universe corresponding to negative or positive value of the 
extrinsic curvature, respectively. 
\par
Hamiltonian constraint \eqref{Hamicon} is invariant under the following scaling with a constant $C$:
\begin{equation}
\Psi\to C\Psi~,~~~~K\to K/C^{2}.
\end{equation}
This ambiguity corresponds to a choice of the unit of the scale. 
In this paper, the normalization is 
fixed by
\begin{equation}
\int_{\rm box} d^{3}x\sqrt{\gamma}\Psi =1.
\label{normalization condition}
\end{equation}
With the conditions \eqref{Kval} and \eqref{normalization condition}, Eq.~\eqref{Hamicon} is numerically integrated by 
using the successive over-relaxation (SOR) method. 
\par
We show that our initial data 
contains gravitational waves in the linear approximation 
$\mathcal{A}\ll1$. 
It is worthwhile to note that, 
since $\tilde R=\mathcal O(\mathcal A^2)$, 
we obtain $K=\mathcal O(\mathcal A)$ from Eq.~\eqref{Kval} and 
$\Psi=1+\mathcal O(\mathcal A^2)$ from 
the Hamiltonian constraint \eqref{Hamicon}. 
In this approximation, the conformal metric becomes 
\begin{equation}
\begin{array}{rcl}
\tilde{\gamma}_{ij}&\simeq&
\delta_{ij}+\mbox{diag}(h^{(3)}-h^{(2)}, h^{(1)}-h^{(3)}, h^{(2)}-h^{(1)}),
\end{array}\label{standing_wave_approx}
\end{equation}
and the fluctuation part $\tilde{\gamma}_{ij}-\delta_{ij}$ satisfies the transverse traceless condition. 
Furthermore, for $\beta=0$, as the equation of motion of $\tilde{\gamma}_{ij}$ is $\frac{\partial}{\partial t}\tilde{\gamma}_{ij}=-2\alpha \tilde{A}_{ij}$, $\tilde{A}_{ij}=0$ means $\frac{\partial}{\partial t}\tilde{\gamma}_{ij}=0$.
Hence, the linearized form of our initial data $\tilde{\gamma}_{ij}$ corresponds to the standing gravitational wave\dred{s} at the moment of maximum amplitude.
\par
Remaning parameter $\mathcal{A}$ corresponds to 
the initial amplitude of gravitational waves in the linear regime. 
In the short wavelength, 
the linear gravitational waves have only 
oscillatory modes 
and their amplitude 
decreases (increases) 
with 
expansion (contraction) 
of the universe. 
On the other hand, 
for the super horizon wavelength, 
the two modes of the gravitational wave\dred{s} 
correspond to the decaying mode and the growing mode. 
The ratio between 
these two modes 
is fixed by the phase of the gravitational waves at the horizon crossing. 
In our initial data, the phase of the gravitational standing waves 
is fixed so that the 
waves have the maximum amplitude at the initial time 
and the change of the value $\mathcal{A}$ causes the change of 
the initial Hubble scale. 
Therefore the phase at the horizon crossing 
depends on $\mathcal{A}$ and  
the behavior of the gravitational waves 
in the long wavelength regime also depends on the initial amplitude $\mathcal{A}$. 
We use initial data with $\mathcal{A}=0.07, 0.08, 0.09, 0.1, 0.11$ 
and analyze the behavior of the metric both in the short wavelength region and long wavelength region. 

\section{Time evolution}\label{Time evolution}
To simulate the time evolution, 
we use the COSMOS code by appropriately modifying it for our purpose. 
The COSMOS code is an Einstein equation solver written in 
C++ by means of BSSN formalism \cite{Shibata:1995we,Baumgarte:1998te}. 
The algorithm of this code is similar to the SACRA code \cite{Yamamoto:2008js}. 
The COSMOS code has been used in papers~\cite{Yoo:2014boa,Yoo:2013yea}. 
In the original code, the 4-th order finite differencing in space with 
uni-grid and the 4th order time integration 
with Runge-Kutta method in Cartesian coordinates are adopted. 
In this paper, we adopt the 2nd order central differencing in space 
and the leapfrog method with a time filtering for time integration. 
Because of the time reversal of the Einstein equation, the time evolution is simulated not only for 
the expanding universe 
with $K<0$, 
but also for the contracting direction from the same initial data with $K>0$. 
Our simulation has been terminated when the violation of the 
Hamiltonian constraint exceeds 1\%.
\subsection{Gauge condition}
Before performing the time integration, we need to fix the gauge conditions 
for the lapse function $\alpha$ and the shift vector $\beta^{i}$. 
In regard to the shift vector, we set $\beta^{i}=0$ 
for simplicity. 
As for the lapse function, we use the ``uniform $K$ gauge" condition 
which keeps spatially uniform $K$ on each time slice. 
\par
Let us derive the equation for the lapse function with the uniform $K$ gauge. 
The time evolution equation of $K$ is given by 
\begin{equation}
\frac{\partial}{\partial t}K=-D_iD^i \alpha
+\alpha\left(\tilde{A}_{ij}\tilde{A}^{ij}+\frac{K^{2}}{3}\right). 
\label{deltK}
\end{equation}
The uniform $K$ gauge condition implies that the  
left hand side in this equation is spatially constant. 
Integrating this equation over the box, 
the time derivative of $K$ is obtained by 
\begin{equation}
\frac{\partial K}{\partial t}=\frac{\int_{\rm box} d^{3}x \sqrt{\gamma}\alpha\left(\tilde{A}_{ij}\tilde{A}^{ij}+\frac{K^{2}}{3}\right)}{\int_{\rm box} d^{3}x\sqrt{\gamma}}, 
\end{equation}
where the first term in the right hand side of Eq.~\eqref{deltK} vanishes 
by virtue of Gauss's theorem 
and the periodic boundary conditions. 
At every time step, we solve Eq.~\eqref{deltK} by the SOR method to determined $\alpha$. 
\subsection{Physical quantities}\label{Sec Physical quantities}
In order to test Isaacson's formula (\ref{Isaacson formula}), 
we investigate the relation between the effective Hubble parameter 
and the effective scale factor. 
Since 
our model is inhomogeneous, 
there is no unique definition of the Hubble parameter and the scale factor. 
Nevertheless, by virtue of the uniform $K$ gauge condition, 
the Hubble parameter can be naturally defined as 
\begin{equation}
H\equiv -\frac{K}{3}. 
\end{equation}
This definition 
coincides with 
the standard definition of 
$H$ for the homogeneous and isotropic universe model. 
\par
\begin{figure}
\centering{
\includegraphics[width=7cm]{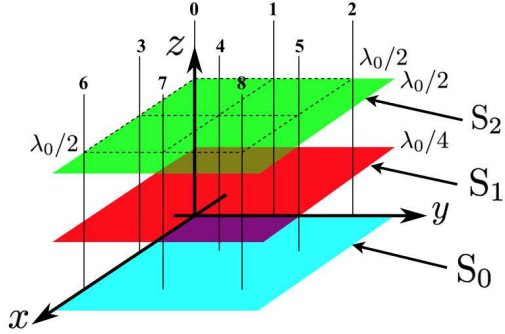}
}
\caption{
The box 
with periodic boundaries. 
The definitions of the lines 0-8~
and surfaces 
${\rm S}_{{\rm 0}}$($z=0$, blue surface), ${\rm S}_{{\rm 1}}$($z=\lambda_{0}/4$, red surface) and ${\rm S}_{{\rm 2}}$($z=\lambda_{0}/2$, green surface) are 
shown.
}\label{periodic_box}
\end{figure}

One of the simplest definition of the scale factor is defined from the proper volume of the box: 
\begin{equation}\label{the definition of scale factor from the proper volume}
L_{\rm V}\equiv \left(\int_{\rm box} dxdydz\sqrt{\gamma}\right)^{1/3}.
\end{equation}
We can also define other effective scale factors 
from the edge proper length and the surface proper area. 
These definitions are used in the several previous 
works\cite{Bentivegna:2012ei,Yoo:2012jz,Yoo:2014boa,Yoo:2013yea,Clifton:2009jw,Clifton:2012qh}. 
The effective scale factors from the proper length and the proper area are 
\begin{eqnarray}\label{the definition of scale factor from the proper length}
L_{\rm L}(x,y)&\equiv&\int^{\lambda_{0}}_{0} dz \sqrt{\gamma_{zz}}, \\
\label{the definition of scale factor from the proper area}
L_{\rm A}(z)&\equiv&\left(\int^{\lambda_{0}}_{0} dxdy\sqrt{\gamma_{xx}\gamma_{yy}
-\gamma_{xy}^{2}}\right)^{1/2}. 
\end{eqnarray}
Since $L_{\rm L}$ and $L_{\rm A}$ depend on $x,y,z$,
we pick up the several characteristic positions as shown in Fig.~\ref{periodic_box}. 
$L_{\rm A}(z)$ is evaluated at $z=0,\lambda_{0}/4,\lambda_{0}/2$ which are labeled as ${\rm S}_{{\rm 0}}$, ${\rm S}_{{\rm 1}}$ and ${\rm S}_{{\rm 2}}$, respectively. 
$L_{\rm L}(x,y)$ is evaluated at 
$(x,y)=(0,0)$, $(0,\lambda_{0}/4)$,$(0,\lambda_{0}/2)$,$(\lambda_{0}/4,0)$ 
,$(\lambda_{0}/4,\lambda_{0}/4)$,$(\lambda_{0}/4,\lambda_{0}/2)$,$(\lambda_{0}/2,0)$
,$(\lambda_{0}/2,\lambda_{0}/4)$,\\$(\lambda_{0}/2,\lambda_{0}/2)$, 
which are labeled as 0-8 lines, respectively. 
\par
In order to compare our numerical results with Isaacson's formula~\eqref{Isaacson formula} 
in a precise sense, 
we must take not only spatial average but also temporal average. 
We consider the spatial average is taken by the definition of effective scale factors 
~\eqref{the definition of scale factor from the proper volume}-\eqref{the definition of scale factor from the proper area}. 
However, for temporal average, we do not have any appropriate 
scale for the average, since our model has no exact periodicity 
along the time direction. 
One possible way for the temporal average is proposed and 
discussed in the Sec.\ref{sec temporal average}. 
\par
We evaluate the difference between 
the effective scale factor from 
our numerical result and 
that of 
Isaacson's formula 
as a function of the Hubble parameter. 
The deviation is evaluated by 
\begin{equation}
\delta_{\rm V}(H)\equiv\frac{L_{\rm V}-L^{\rm Isa}_{\rm V}}{L^{\rm Isa}_{\rm V}},
\label{eqdv}
\end{equation}
\begin{equation}
\delta_{\rm A}(H,z)\equiv\frac{L_{\rm A}-L^{\rm Isa}_{\rm A}}{L^{\rm Isa}_{\rm A}},
\label{eqda}
\end{equation}
\begin{equation}
\delta_{\rm L}(H,x,y)\equiv\frac{L_{\rm L}-L^{\rm Isa}_{\rm L}}{L^{\rm Isa}_{\rm L}},
\label{eqdl}
\end{equation}
where  $L^{\rm Isa}_{\rm V}(H)$, $L^{\rm Isa}_{\rm A}(H,z)$ and $L^{\rm Isa}_{\rm L}(H,x,y)$ 
represent Isaacson's formula~\eqref{Isaacson formula} 
and the coefficient $b$ is determined by the least squares fitting of our numerical result 
in the region $L_{\rm V}<\lambda_{0}$, which is attained by time evolution along the expanding temporal direction. 

\subsection{Convergence check}
\begin{figure}
\center{
\includegraphics[width=9cm]{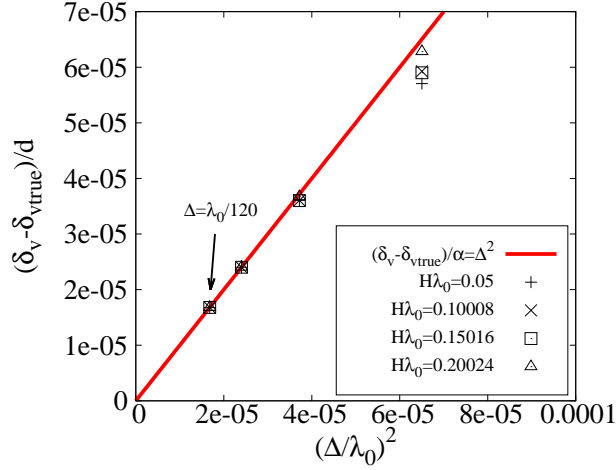}
}
\caption{
The convergence of $(\delta_{\rm V}(\Delta)-\delta_{\rm{Vtrue}})/d$ at selected time steps. 
$\delta_{\rm Vtrue}$ and $d$ are fixed by Eqs.~\eqref{alpha} and \eqref{vtrue}. 
}
\label{convergence-Lv}
\end{figure}

Let us consider the value of $\delta_{\rm V}$ as a function of 
the grid spacing $\Delta$. 
Since our numerical code has the second order accuracy, 
$\delta_{V}(\Delta)$ and its true value $\delta_{\rm V true}\equiv\delta_{\rm V}(0)$ 
are supposed to have the relation 
\begin{equation}
\delta_{\rm V}(\Delta)=\delta_{\rm{V true}}+d \Delta^{2}
+\mathcal O(\Delta^3). 
\end{equation}
where a coefficient $d$ is determined by numerical results. 
Using the two data sets taken with $\Delta=\Delta_{1}$ and $\Delta_{2}$, 
$\delta_{\rm{Vtrue}}$ and $d$ can be 
obtained as 
\begin{eqnarray}
d&=&\frac{\delta_{\rm V}(\Delta_{1})-\delta_{\rm V}(\Delta_{2})}{\Delta_{1}^{2}-\Delta_{2}^{2}}+\mathcal{O}(\Delta^3),
\label{alpha}
\\
\delta_{\rm V true}&=&\frac{\delta_{\rm V}(\Delta_{2})\Delta_{1}^{2}-\delta_{\rm V}(\Delta_{1})\Delta_{2}^{2}}{\Delta_{1}^{2}-\Delta_{2}^{2}}+\mathcal{O}(\Delta^3)
\label{vtrue}
\end{eqnarray}
For $\Delta_1=\lambda_{0}/120$ and  $\Delta_2=\lambda_{0}/100$, 
we calculate the value of $d$ and $\delta_{\rm V true}$. 
The value of convergence $(\delta_{\rm V}(\Delta)-\delta_{\rm V true})/d$ is 
compared with $\Delta^2$ for each value of $\Delta$ in Fig.~\ref{convergence-Lv}. 
We can clearly confirm that the second order convergence is achieved from Fig.~\ref{convergence-Lv}. 
In the following, we use the numerical result with $\Delta=\lambda_{0}/120$. 

\section{Results of simulation}
\subsection{Relation between Hubble parameter and effective scale factor}
In this subsection, we explain the relation between the Hubble parameter and 
the effective scale factors
$L_{\rm V}$, $L_{\rm A}$, $L_{\rm L}$ 
for the initial amplitude $\mathcal{A}=0.1$. 
\par
Before discussing the result, 
we introduce the characteristic Hubble parameter 
$H_{1.0}$, $H_{0.5}$ and $H_{0.1}$ by 
\begin{equation}
L^{\rm Isa}(H_{1.0})=H^{-1},~~L^{\rm Isa}(H_{0.5})=0.5H^{-1},~~L^{\rm Isa}(H_{0.1})=0.1H^{-1},
\end{equation}
 where $L^{\rm Isa}(H)=\sqrt{b /H}$. 
That is, $H^{-1}=H^{-1}_{1.0}$ corresponds to the horizon crossing Hubble time. 
The parameter $b$ in each $L^{\rm Isa}(H)$ is determined by fitting the numerical data obtained by time evolution in expanding temporal direction. 
We define the short wavelength region $L^{\rm Isa}<H^{-1}$ and the long wavelength region $L^{\rm Isa}>H^{-1}$ in comparison with the Hubble scale. 
\par
\begin{figure}[H]
\begin{tabular}{cc}
\includegraphics[width=7cm]{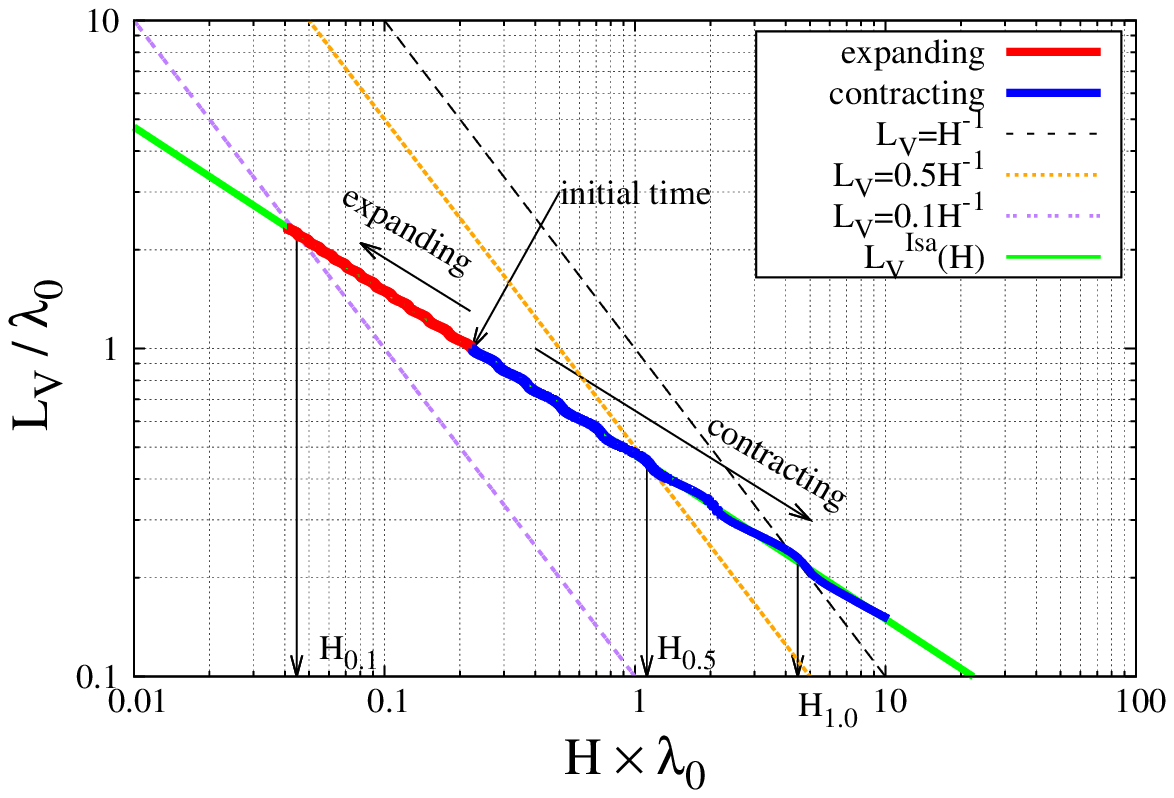}&
\includegraphics[width=7cm]{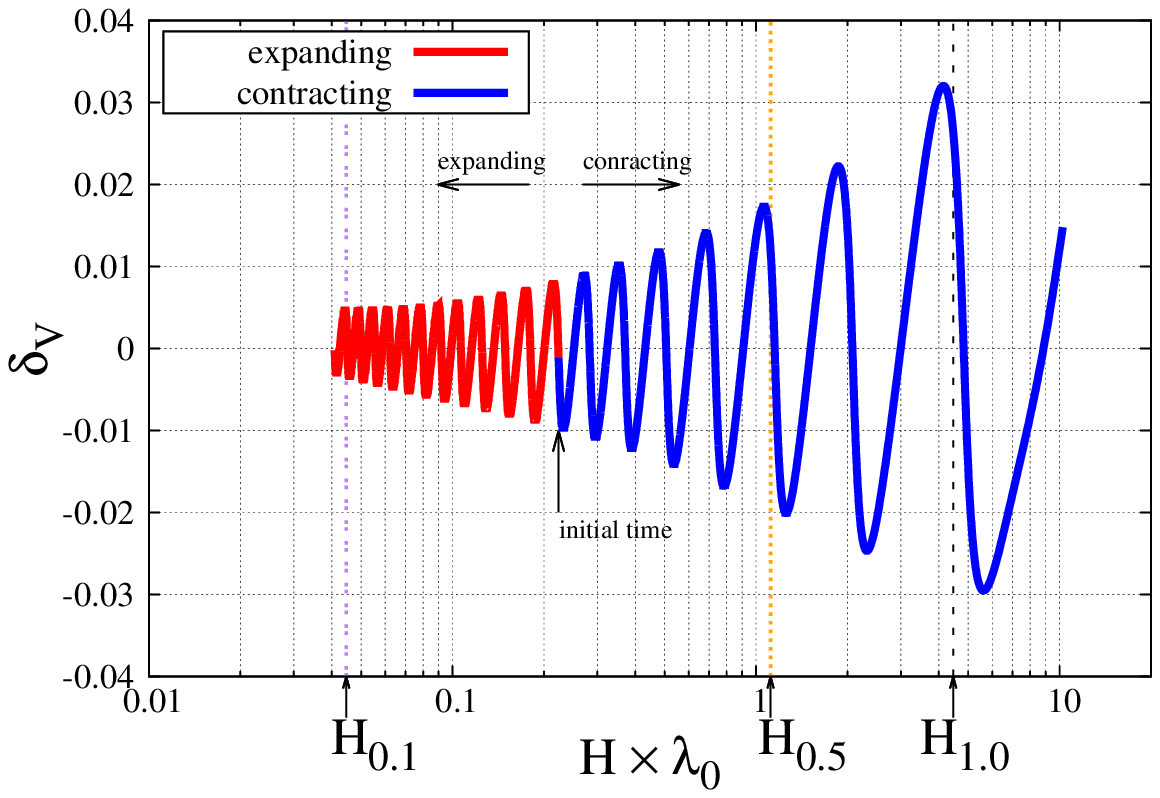}\\
\end{tabular}
\caption{
The behavior of $L_{\rm V}(H)$. 
Isaacson's formula is $L^{\rm Isa}_{\rm V}(H)=\sqrt{b_{\rm V}/H}$ with $b_{\rm V}=0.224$. 
}\label{the relation of H and LV}
\end{figure}

The behavior of $L_{\rm V}(H)$ and $\delta_{\rm V}(H)$ is shown in Fig.~\ref{the relation of H and LV}. 
\noindent
These figures show 
the oscillation of the scale factor \magenta{which} 
reflects the \magenta{oscillation of} gravitational waves. 
Our simulations have been terminated when the proper wave length 
$L_{\rm V}$ 
exceeds about $1.5/H$ in the contracting direction. 
It should be noted that Isaacson's formula 
is not guaranteed if the proper wave length becomes comparable to the Hubble scale. 
Nevertheless, Fig.~\ref{the relation of H and LV} shows that the 
maximum value of the deviation from Isaacson's formula is about 3\% 
and the deviation of central value from the formula is even smaller. 
Next, we show the behavior of $L_{\rm A}(H)$ and $L_{\rm L}(H)$. 
The behavior of 
$L_{\rm A0}(H)$(${\rm S}_{{\rm 0}}$ in Fig.~\ref{periodic_box}), $L_{\rm A1}(H)$ (${\rm S}_{{\rm 1}}$ in Fig.~\ref{periodic_box}) and 
$L_{\rm A2}(H)$(${\rm S}_{{\rm 2}}$ in Fig.~\ref{periodic_box}) 
are almost same as each other.
Therefore, we plot only $L_{\rm A0}(H)$. 
\begin{figure}[H]
\begin{tabular}{cc}
\includegraphics[width=7cm]{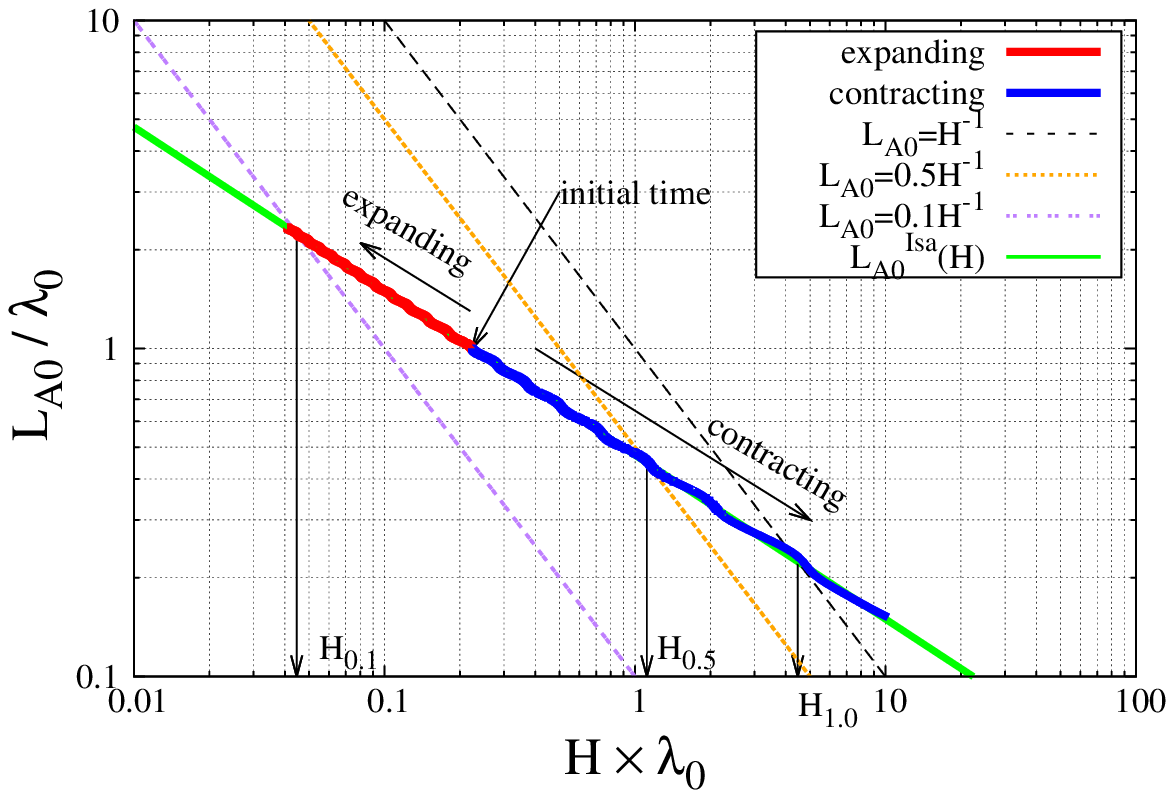}&
\includegraphics[width=7cm]{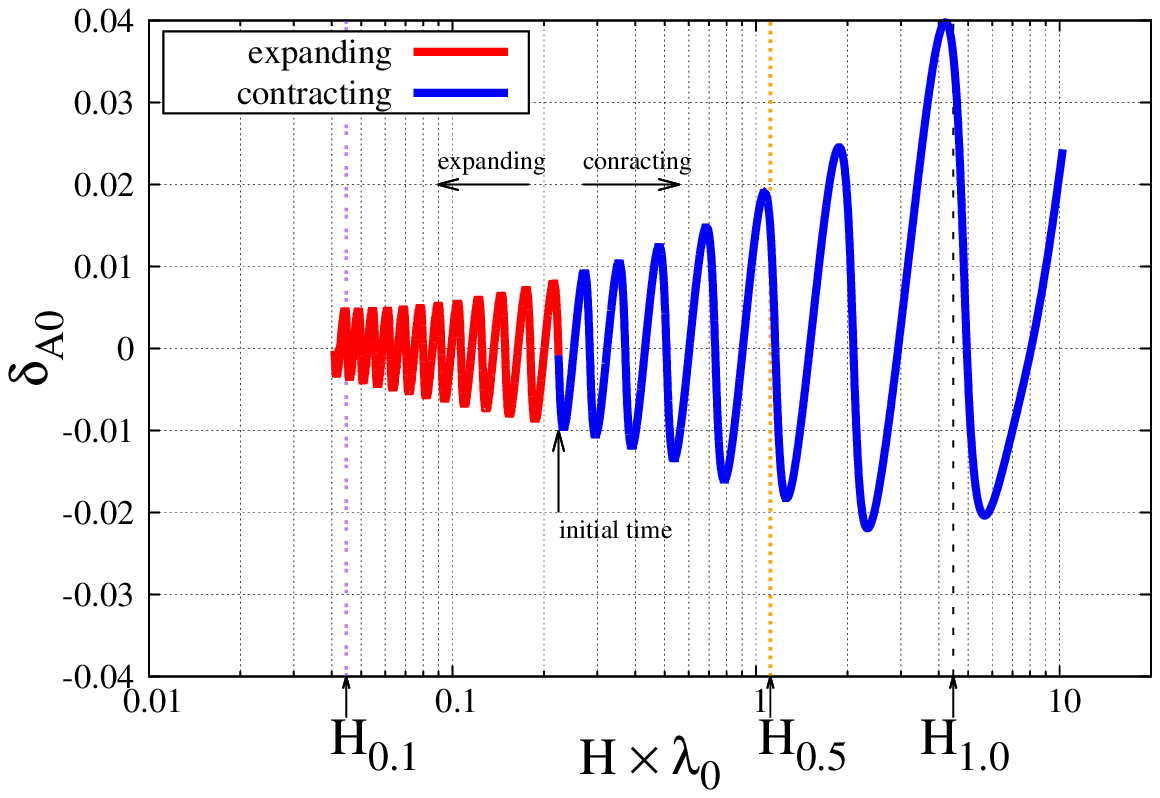}\\
\end{tabular}
\caption{
The behavior of $L_{\rm A0}(H)$. 
Isaacson's formula is $L^{\rm Isa}_{\rm A0}(H)=\sqrt{b_{\rm A0}/H}$ with $b_{\rm A0}=0.224$. 
}\label{the relation of H and LA0}
\end{figure}
\noindent
As is shown in Fig.~\ref{the relation of H and LA0}, 
$L_{\rm A0}(H)$ also oscillates in a similar way as $L_{\rm V}(H)$.
The maximum value of the deviation from Isaacson's formula around $H\sim H_{1.0}$ is about $4\%$, 
which is slightly larger than that \dred{of} $L_{\rm V}(H)$.
\par
\begin{figure}[p]
\begin{tabular}{cc}
\includegraphics[width=7cm]{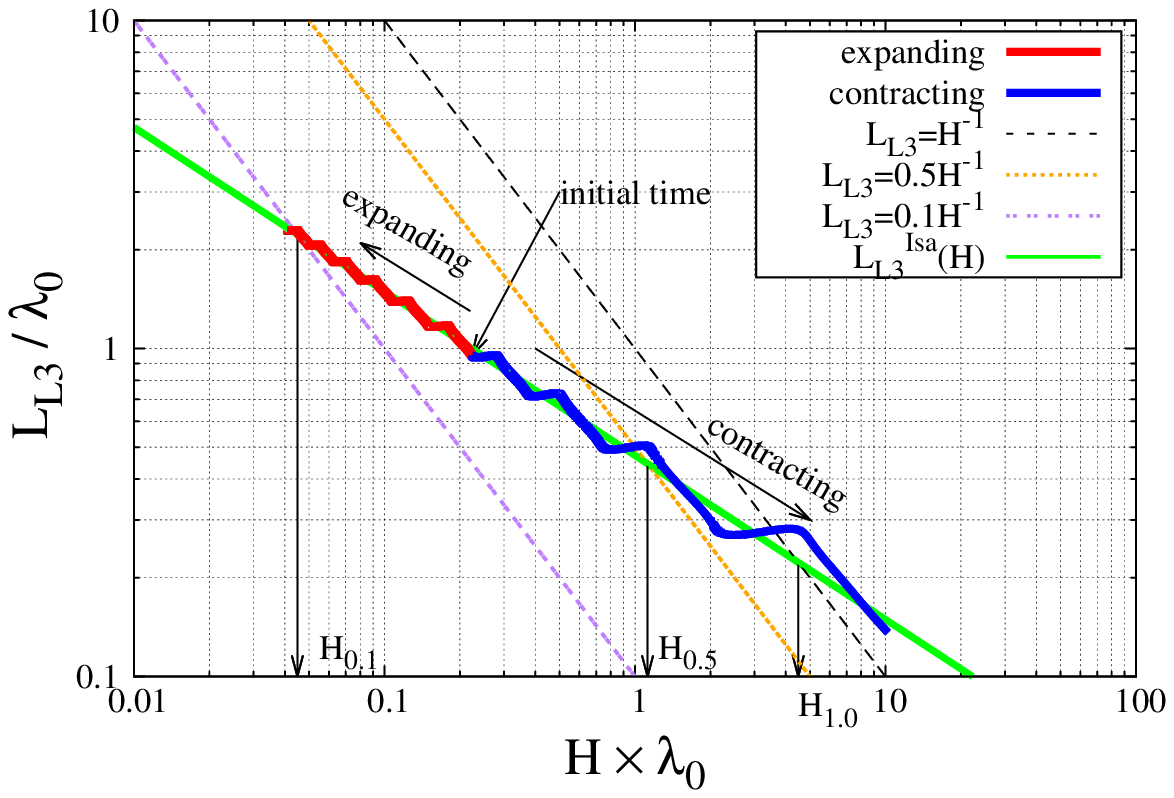}&
\includegraphics[width=7cm]{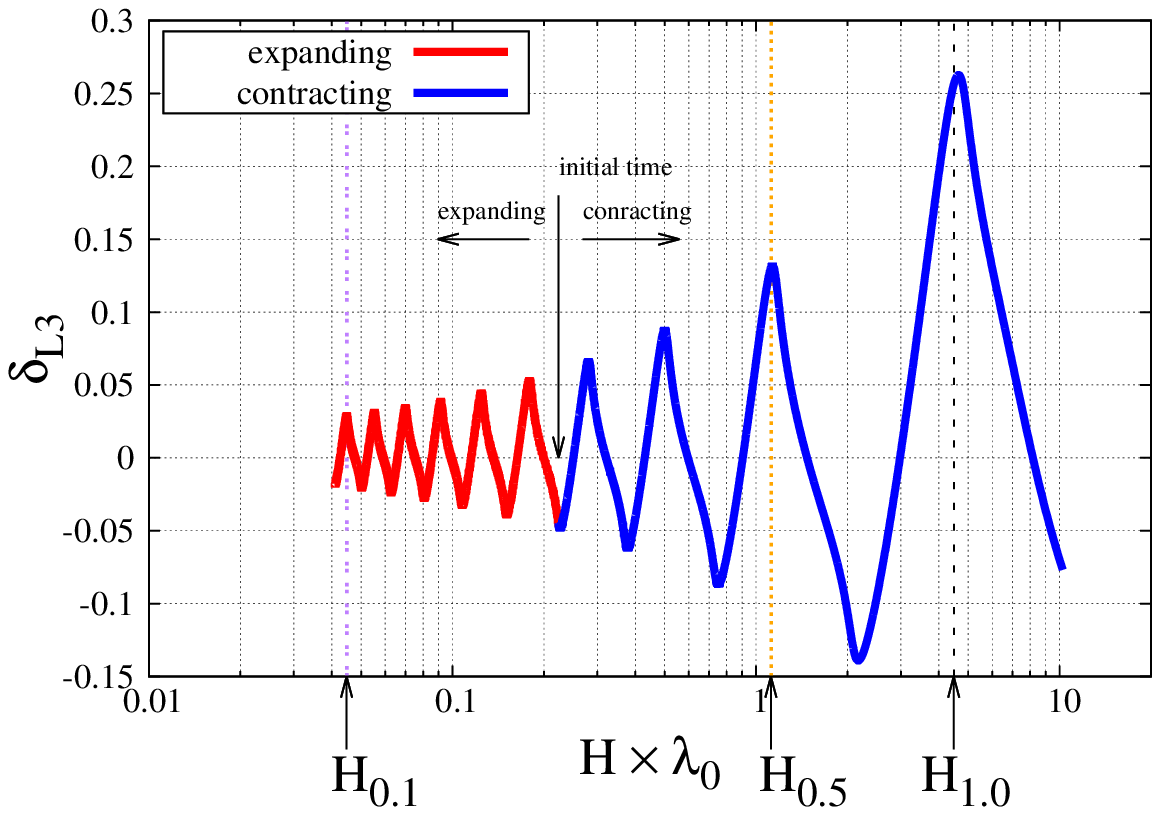}\\
\end{tabular}
\begin{center} {\footnotesize 
$L_{L3}(H)$:~Isaacson's formula is $L_{L3}^{\rm Isa}(H)=\sqrt{b_{L3}/H}$ with $b_{L3}=0.223$.} \end{center}
\begin{tabular}{cc}
\includegraphics[width=7cm]{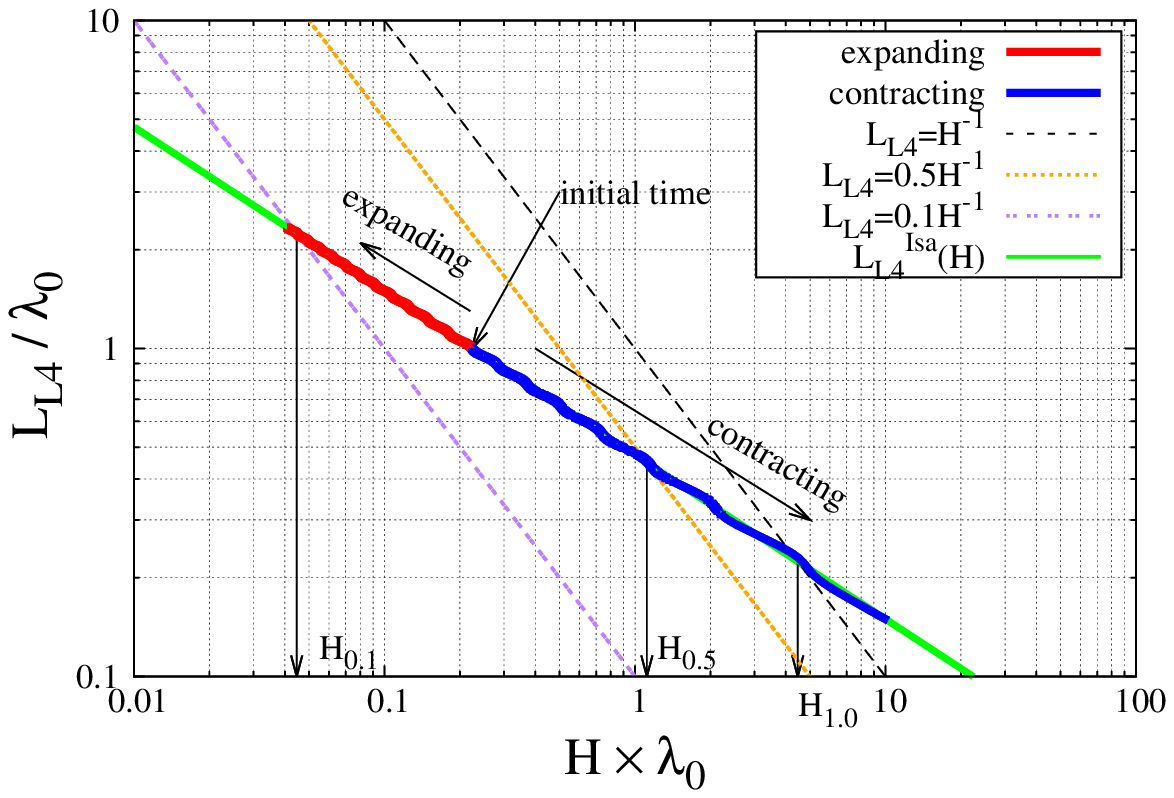}&
\includegraphics[width=7cm]{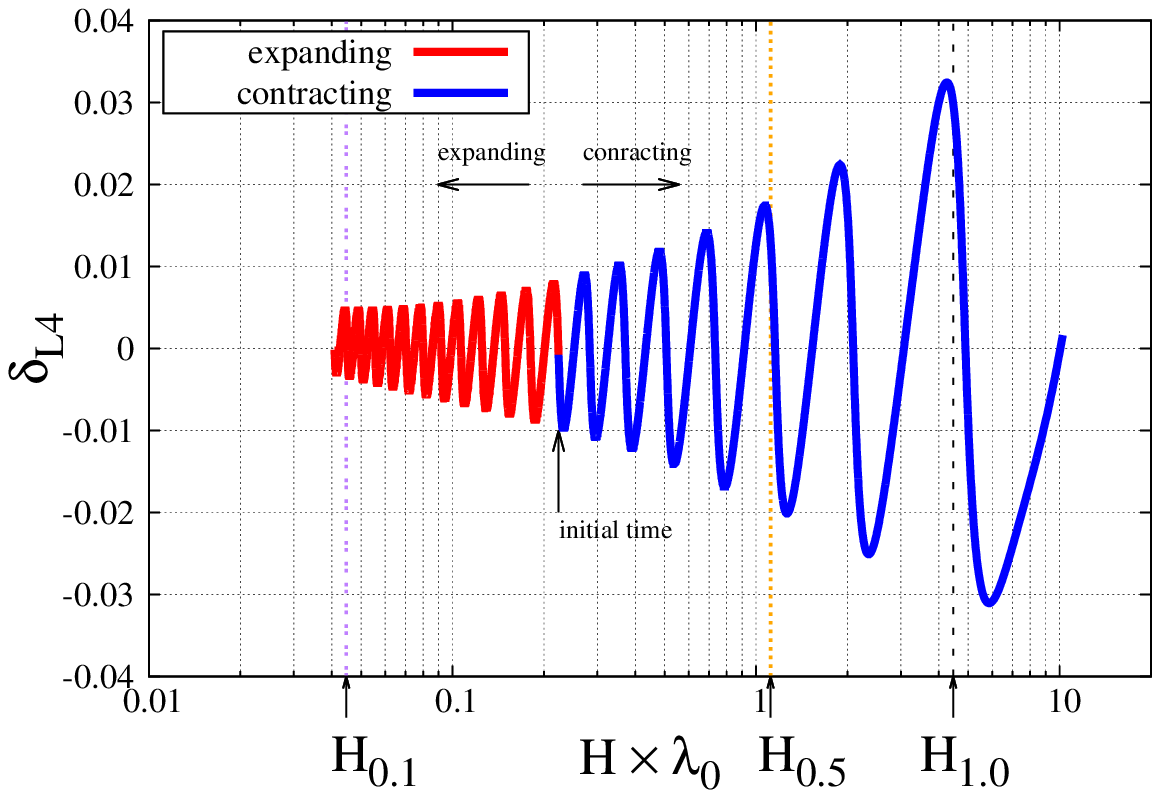}\\
\end{tabular}
\begin{center} {\footnotesize 
$L_{L4}(H)$:~Isaacson's formula is $L_{L4}^{\rm Isa}(H)=\sqrt{b_{L4}/H}$ with $b_{L4}=0.224$.} \end{center}
\begin{tabular}{cc}
\includegraphics[width=7cm]{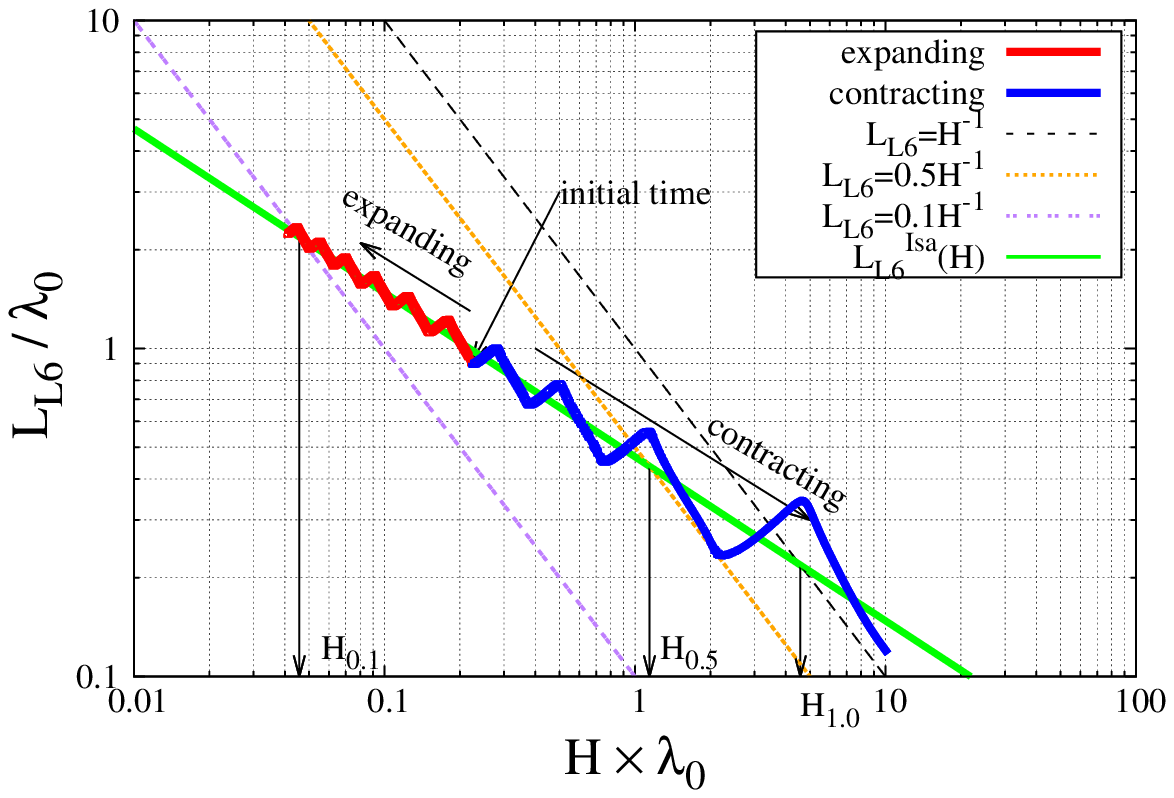}&
\includegraphics[width=7cm]{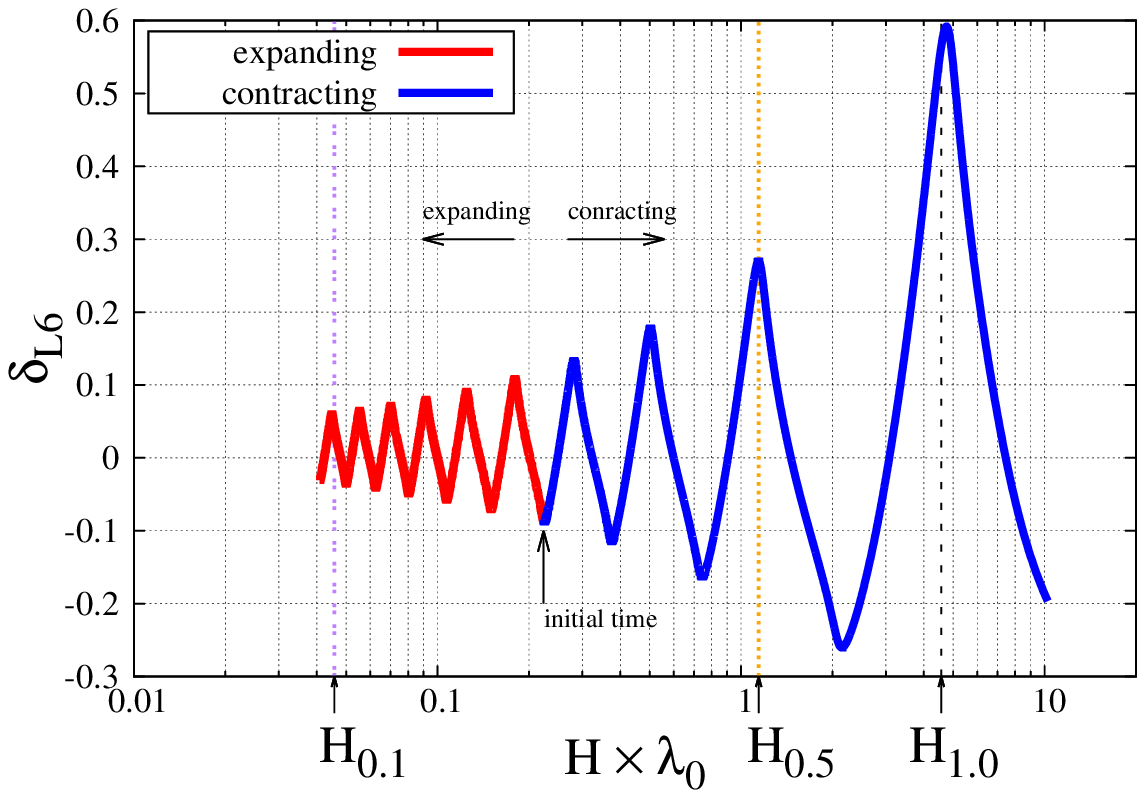}\\
\end{tabular}
\begin{center} {\footnotesize 
$L_{L6}(H)$:~Isaacson's formula is $L_{L6}^{\rm Isa}(H)=\sqrt{b_{L6}/H}$ with $b_{L6}=0.218$.} \end{center}
\caption{
The characteristic behavior of $L_{L}(H)$ and $\delta_{L}(H)$.  
}\label{the relation of H and L}
\end{figure}

The relation $L_{\rm L}(H)$
depends on the line position and can be classified into six types (Fig.~\ref{the relation of H and L}). 
\begin{figure}[p]
\begin{tabular}{cc}
\includegraphics[width=7cm]{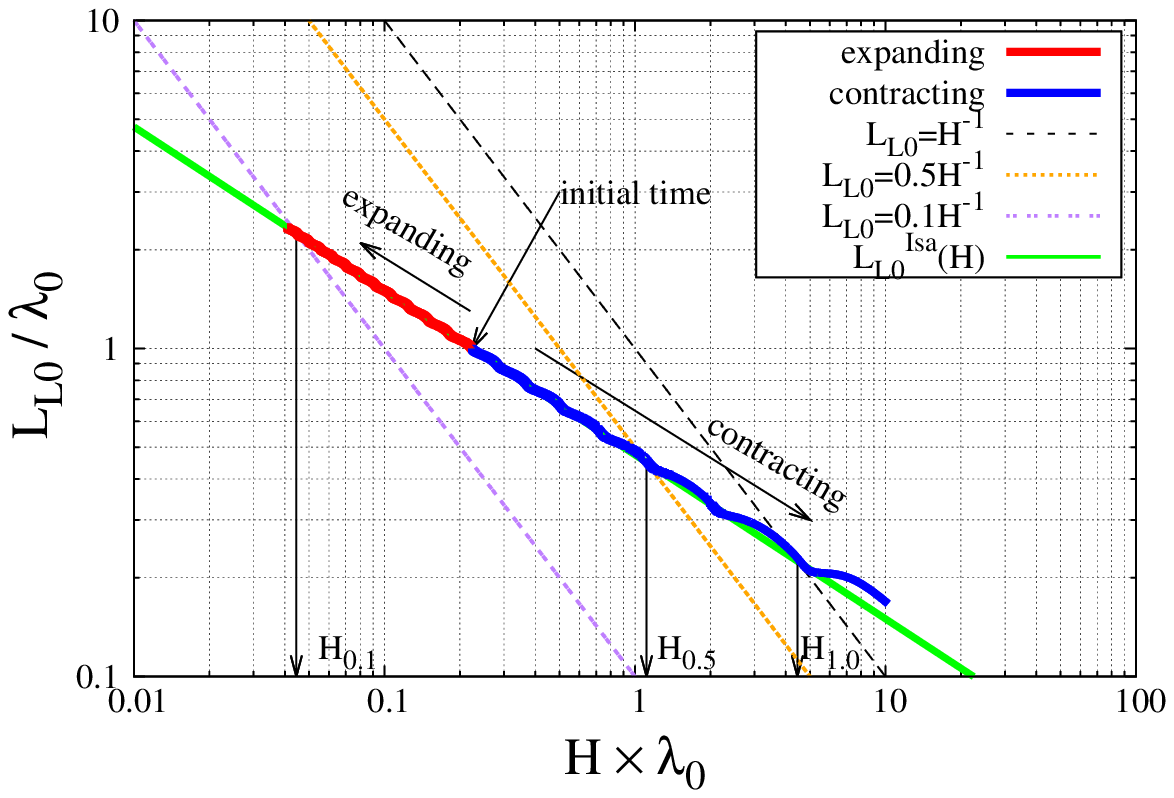}&
\includegraphics[width=7cm]{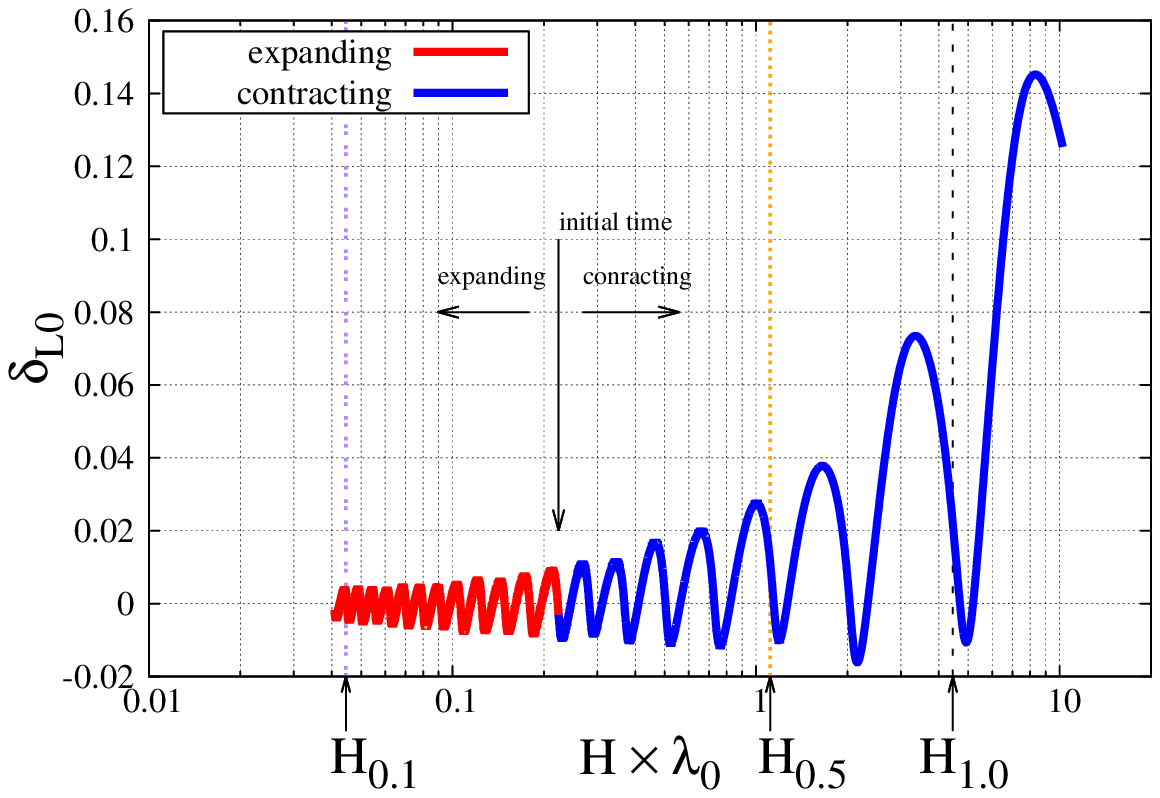}\\
\end{tabular}
\begin{center} {\footnotesize
$L_{L0}(H)$:~Isaacson's formula is $L_{L0}^{\rm Isa}(H)=\sqrt{b_{L0}/H}$ with $b_{L0}=0.224$.} \end{center}
\begin{tabular}{cc}
\includegraphics[width=7cm]{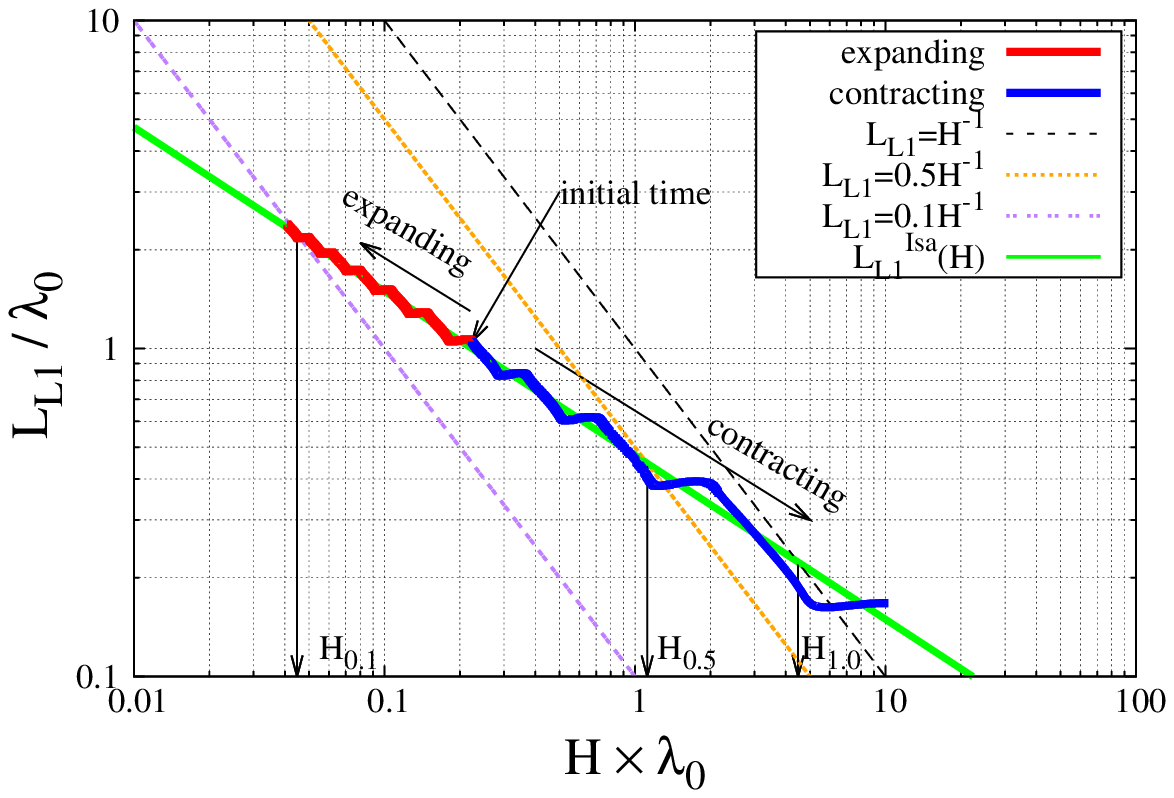}&
\includegraphics[width=7cm]{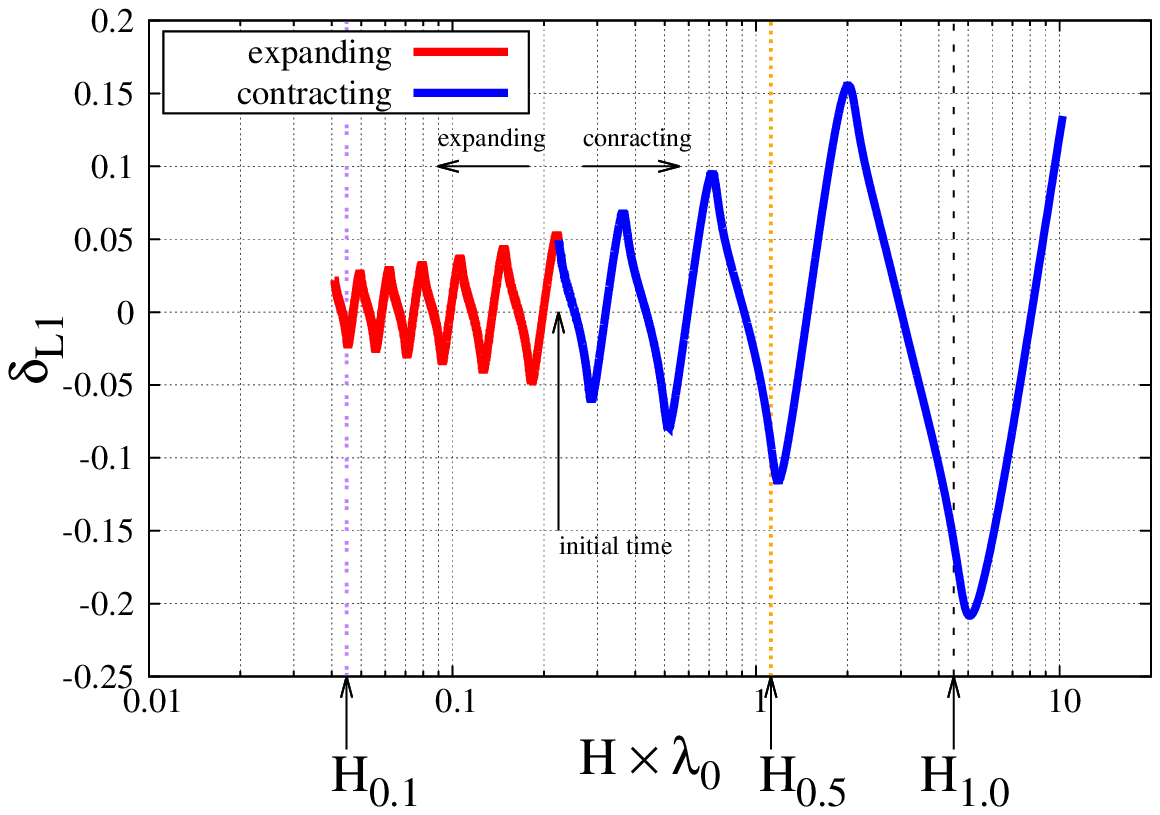}\\
\end{tabular}
\begin{center} {\footnotesize 
$L_{L1}(H)$:~Isaacson's formula is $L_{L1}^{\rm Isa}(H)=\sqrt{b_{L1}/H}$ with $b_{L1}=0.223$.} \end{center}
\begin{tabular}{cc}
\includegraphics[width=7cm]{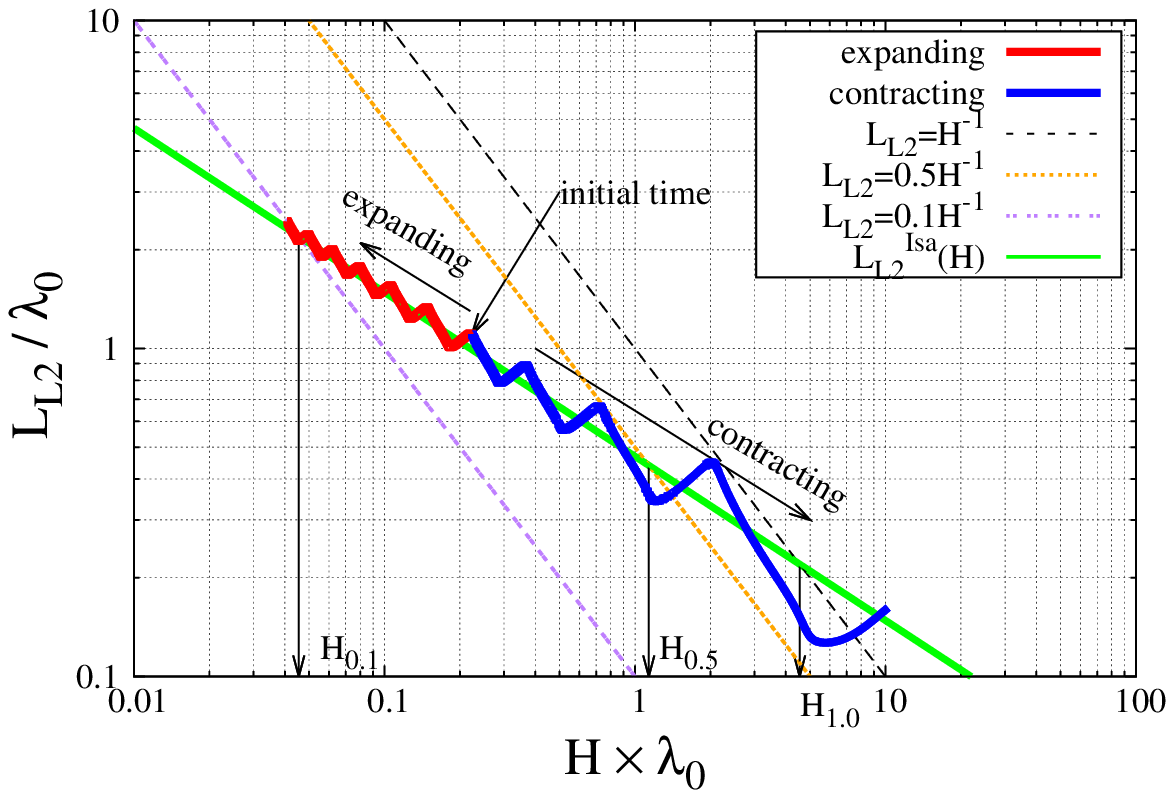}&
\includegraphics[width=7cm]{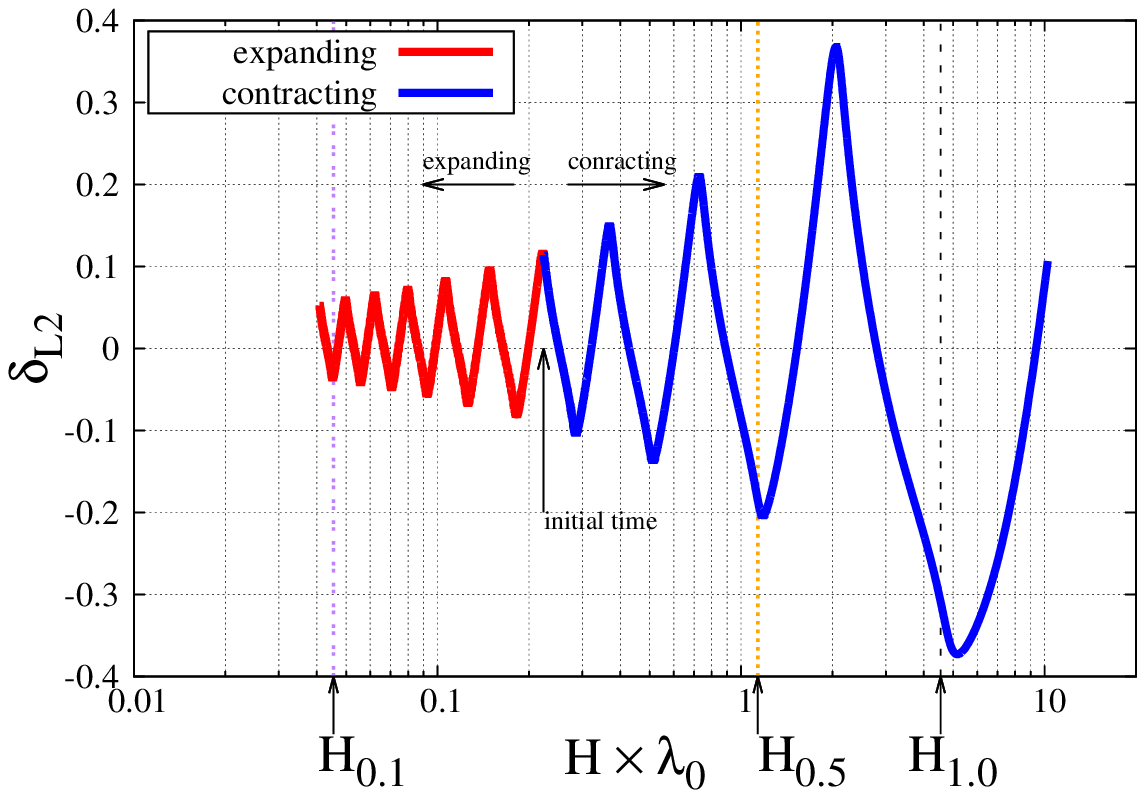}\\
\end{tabular}
\begin{center} {\footnotesize 
$L_{L2}(H)$:~Isaacson's formula is $L_{L2}^{\rm Isa}(H)=\sqrt{b_{L2}/H}$ with $b_{L2}=0.220$.} \end{center}
\end{figure}
\noindent
The behavior of $L_{\rm L8}(H)$ and $\delta_{\rm L8}(H)$ is similar to the behavior of $L_{\rm L0}(H)$ and $\delta_{\rm L0}(H)$. 
Furthermore, the behavior of $\delta_{\rm L1}(H)$ and $\delta_{\rm L3}(H)$ are almost same as $\delta_{\rm L5}(H)$ and $\delta_{\rm L7}(H)$ respectively, 
and the behaviors $\delta_{\rm L2}(H)$ and $\delta_{\rm L6}(H)$ have opposite phase to each other. 
Fig.~\ref{the relation of H and L} shows the value $\max (|\delta_{\rm L0}|: H\sim H_{1.0})=15\%$ and $\max (|\delta_{\rm L4}|: H\sim H_{1.0})=3\%$.
On the other hand, $\max(|\delta_{\rm L2}|:H\sim H_{1.0})=40\%$ and $\max(|\delta_{\rm L6}|:H\sim H_{1.0})=60\%$, which are the maximum in any $\delta_{\rm L}(H\sim H_{1.0})$. 
\par
In order to understand these behaviors and make the position dependence of proper length clear, 
we consider our model with the linear approximation. 
As is mentioned in the Sec.~\ref{Set up of initial data}, 
our model corresponds to the linear standing gravitational waves. 
Since all lines are parallel to $z$-axis, 
only $\gamma_{zz}$ can contribute to the values of proper lengths 
among components of the conformal metric. 
For the standing gravitational waves, 
the time evolution of $\gamma_{zz}$ is given by following form (see from appendix~\ref{Isaacson_in_Freidmann}) 
\begin{equation}\label{behavior of gamma}
\begin{array}{rl}
\gamma_{zz}&\simeq a(t)^{2}\left\{ 1+
 \frac{2\mathcal{A}}{a(t)}\cos \left( \int^{t}_{0}\frac{\omega_{0}dt}{a(t)}+\phi \right) \cos\frac{\pi (x+y)}{\lambda_{0}}\cos\frac{\pi (x-y)}{\lambda_{0}}\right\}, 
\end{array}
\end{equation}
where $\phi$ is a constant determined by the condition $\frac{\partial}{\partial t}\tilde{\gamma}_{ij}|_{t=0}=0$.
\noindent
According to Eq.~(\ref{behavior of gamma}), on the line $0$, $4$ and $8$ 
with coordinates $(x,y)=(0,0)$, $(\frac{\lambda_{0}}{4},\frac{\lambda_{0}}{4})$, $(\frac{\lambda_{0}}{2},\frac{\lambda_{0}}{2})$, $\gamma_{zz}$ does not oscillate in the linear approximation. 
On the other hand, $\gamma_{zz}$ on the line 2 and 6 with coordinates $(0,\frac{\lambda_{0}}{2})$, $(\frac{\lambda_{0}}{2},0)$ oscillate with amplitude $2\mathcal{A}$.
Thus, in our model, the position dependence of the effective scale factor $L$ comes from the difference of 
the amplitude of gravitational waves on each line, 
which is affected by the constructive interference of two oscillation modes  
$h^{(1)}$ and $h^{(2)}$ (see Eq.~(\ref{standing_wave_approx})). 
This result implies that, when we use a proper length as the effective scale factor, position dependence must be carefully treated. 
\subsection{The dependence of the initial amplitude $\mathcal{A}$}\label{dependence of the initial parameter}
As is mentioned in the previous section, for the 
long wavelength ($L>1/H$), 
gravitational waves are superposition of the decaying mode and the growing mode 
and their ratio depends on the phase of gravitational waves at the horizon crossing. 
Due to 
this phase dependence, 
the behavior of gravitational waves depends on $\mathcal{A}$ 
after the horizon crossing. 
In order to see this dependence, we focus on the time evolution of 
the Fourier component of $\gamma_{xx}$ with the wave number 
$\vec{k}=(2\pi n_{x}/\lambda_{0},2\pi n_{y}/\lambda_{0},2\pi n_{z}/\lambda_{0})$: 
\begin{equation}\label{definition of h}
\gamma_{xx}(t,\vec{k})\equiv\frac{1}{\lambda_{0}^{3}}\int^{\lambda_{0}}_{0}d^{3}x \gamma_{xx}(t,\vec{x})\cos (\vec{k}\cdot\vec{x} ).
\end{equation}
For $\vec{k}_{0}\equiv (0,0,2\pi /\lambda_{0})$, the initial value of $\gamma_{xx}(t,\vec{k}_{0})$ is given by 
\begin{equation}
\gamma_{xx}(0,\vec{k}_{0})=\frac{\mathcal{A}}{2}+\mathcal O(\mathcal A^2).
\end{equation}
\par
On the flat FLRW metric background, 
we can analytically calculate the behavior of linear gravitational waves.
The metric can be expressed as follows: 
\begin{equation}
ds^2=-dt^2+a(t)^2\left\{\delta_{ij}+h_{ij}(t,\vec{x})\right\}dx^idx^j, 
\end{equation}
where $a(t)$ is the scale factor of the background universe 
and $h_{ij}(t,\vec{x})$ is the transverse traceless tensor of gravitational waves in the linear approximation. 
The evolution equation for a Fourier mode 
$h_{ij}(t,\vec{k})$ 
of 
$h_{ij}(t,\vec{x})$  
is given by (see Appendix~\ref{Isaacson_in_Freidmann}) 
\begin{equation}
\frac{\partial^2}{\partial t^2} h_{ij}(t,\vec{k})+3H\frac{\partial}{\partial t} h_{ij}(t,\vec{k})+\frac{1}{a^2}k^2h_{ij}(t,\vec{k})=0. 
\end{equation} 
Behavior of $h_{ij}(t,\vec{k}_{0})$ 
for the short wavelength \magenta{$k_{0}>aH$} can be approximated 
by the WKB form as follows: 
\begin{equation}\label{Dij}
h_{ij}(t,\vec{k}_{0})\simeq\frac{1}{a}h_{ij}|_{t=0}\cos\left(\int^t_0 \frac{k_{0}dt}{a}\right),  
\end{equation}
where we have used the normalization $a|_{t=0}=1$. 
Let us assume that the time evolution of the background metric is given by 
that of radiation dominated universe, 
\begin{equation}\label{scale factor}
a=\sqrt{ 2b  t+1}, 
\end{equation}
where $b=b_{\rm V}$ and $a=L_{\rm V}/\lambda_{0}$. 
From Eq.~(\ref{Dij}), the time evolution of the Fourier component  
is given by 
\begin{equation}
a(t)^{2}h_{xx}(t,\vec{k}_{0})=
a\frac{\mathcal{A}}{2}
\cos \left[-\frac{1}{b\lambda_{0}}\left(\sqrt{2b t+1}-\phi \right)\right],
\label{eqht}
\end{equation}
where we have set $a^{2}h_{xx}|_{t=0}=\gamma_{xx}|_{t=0}$ up to the leading order 
and $\phi$ is the integration constant. 
\par
We want to evaluate the deviation between $\gamma_{xx}(t,\vec{k}_{0})$ and $a(t)^{2}h_{xx}(t,\vec{k}_{0})$. 
Since we have used the WKB and the linear approximation 
to derive the expression \eqref{eqht}, 
when one of these approximations is violated, $a(t)^{2}h_{xx}(t,\vec{k}_{0})$ deviates 
from $\gamma_{xx}(t,\vec{k}_{0})$. 
In our setting, this deviation may happen around the horizon crossing. 
After the horizon crossing 
even in linear regime, the ``decaying mode'' (for expanding universe) 
is enhanced and the deviation of our numerical solution from the linear WKB solution is expected. 
This deviation due to the decaying mode depends on 
the value of $\mathcal A$. 
\par
\begin{figure}[p]
\begin{tabular}{cc}
\includegraphics[width=7cm]{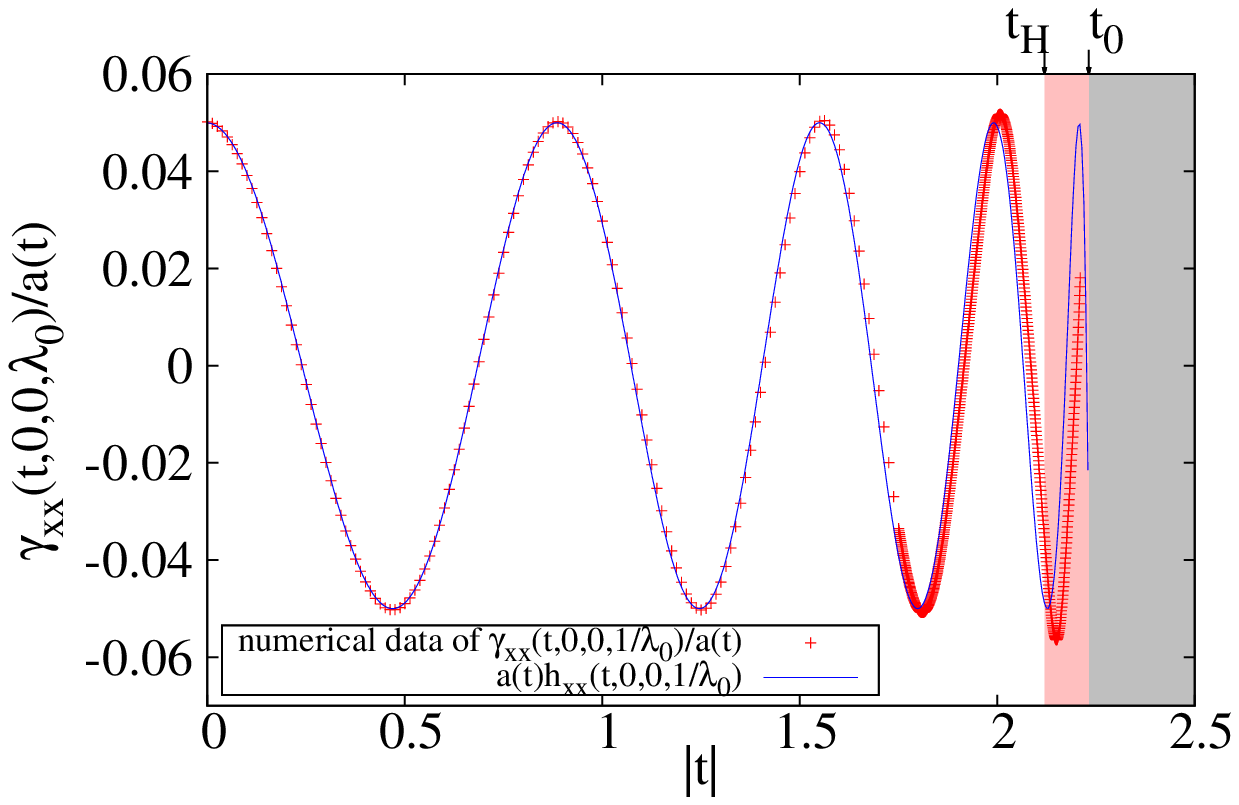}&
\includegraphics[width=7cm]{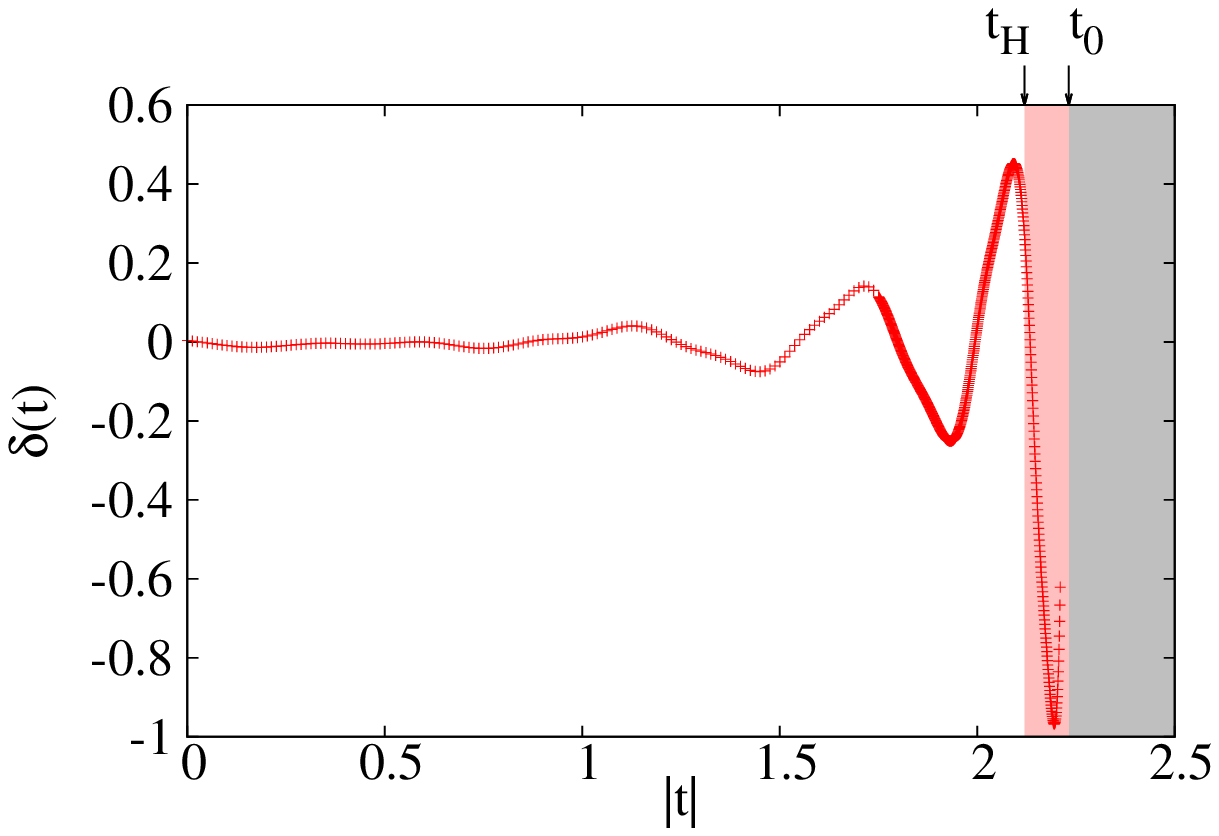}
\end{tabular}
\centering
{\footnotesize $\mathcal{A}=0.1$:$\phi=1.95$.}
\begin{tabular}{cc}
\includegraphics[width=7cm]{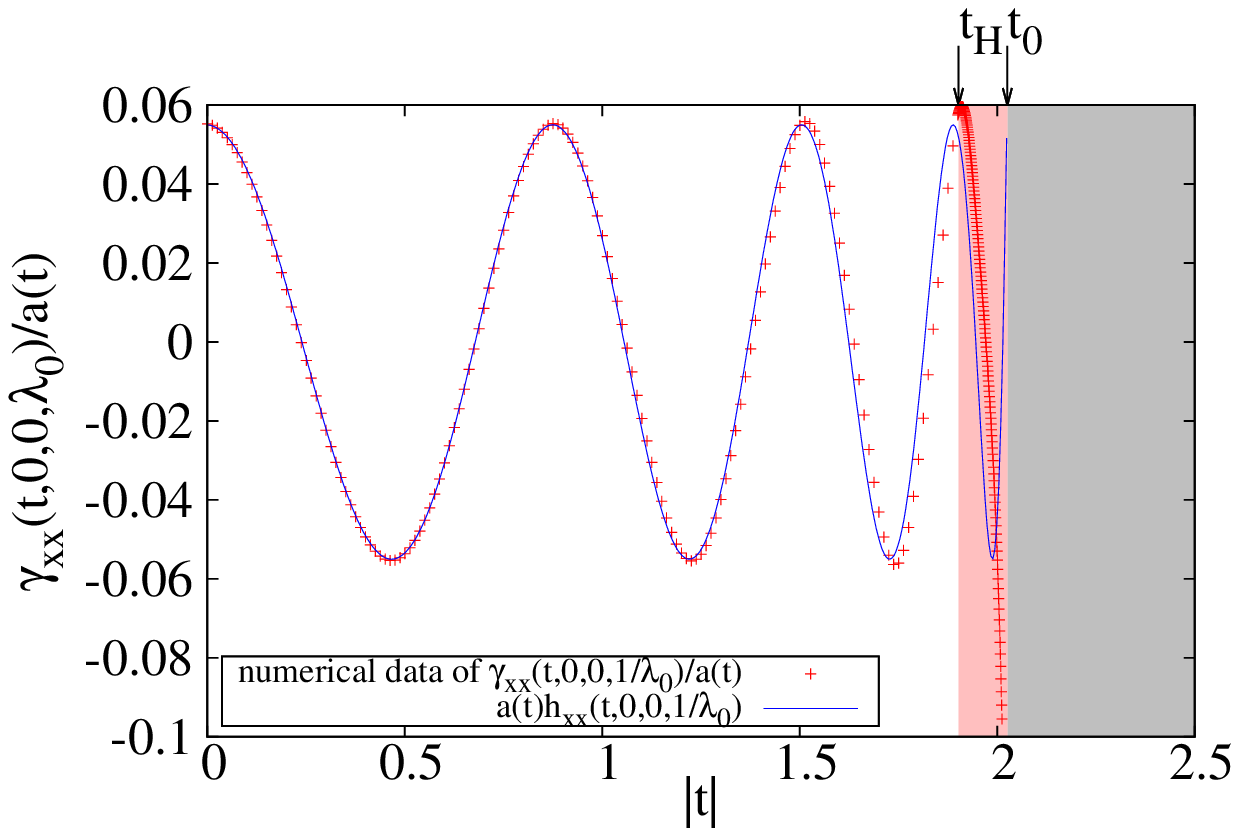}&
\includegraphics[width=7cm]{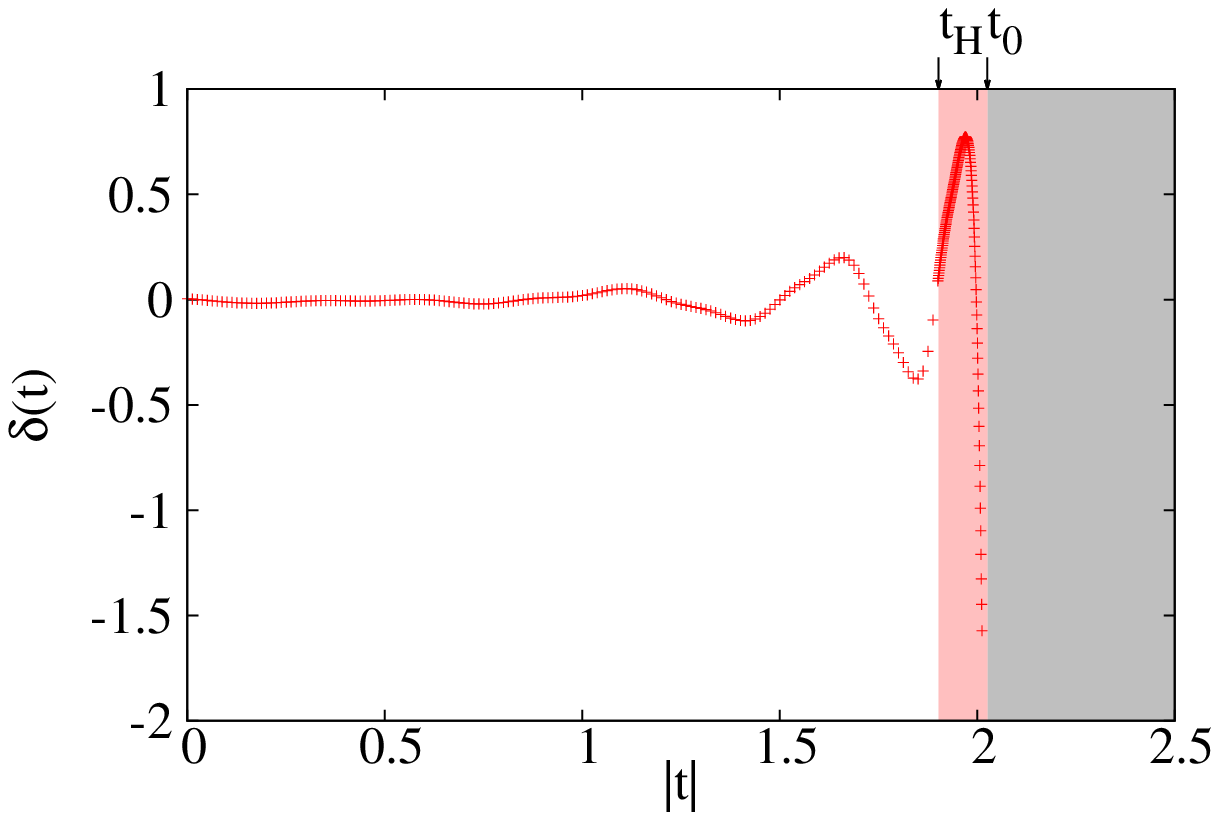}
\end{tabular}
\centering
{\footnotesize $\mathcal{A}=0.11$:$\phi=1.06$.}
\begin{tabular}{cc}
\includegraphics[width=7cm]{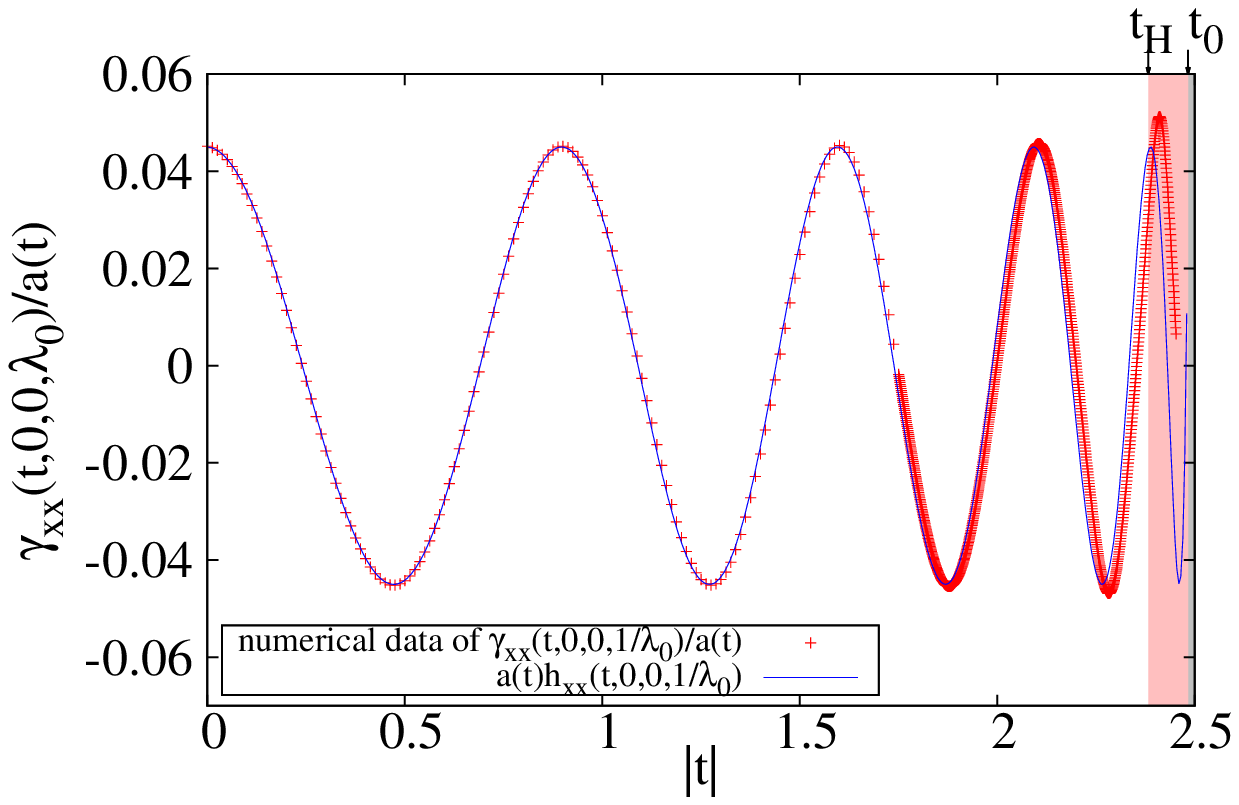}&
\includegraphics[width=7cm]{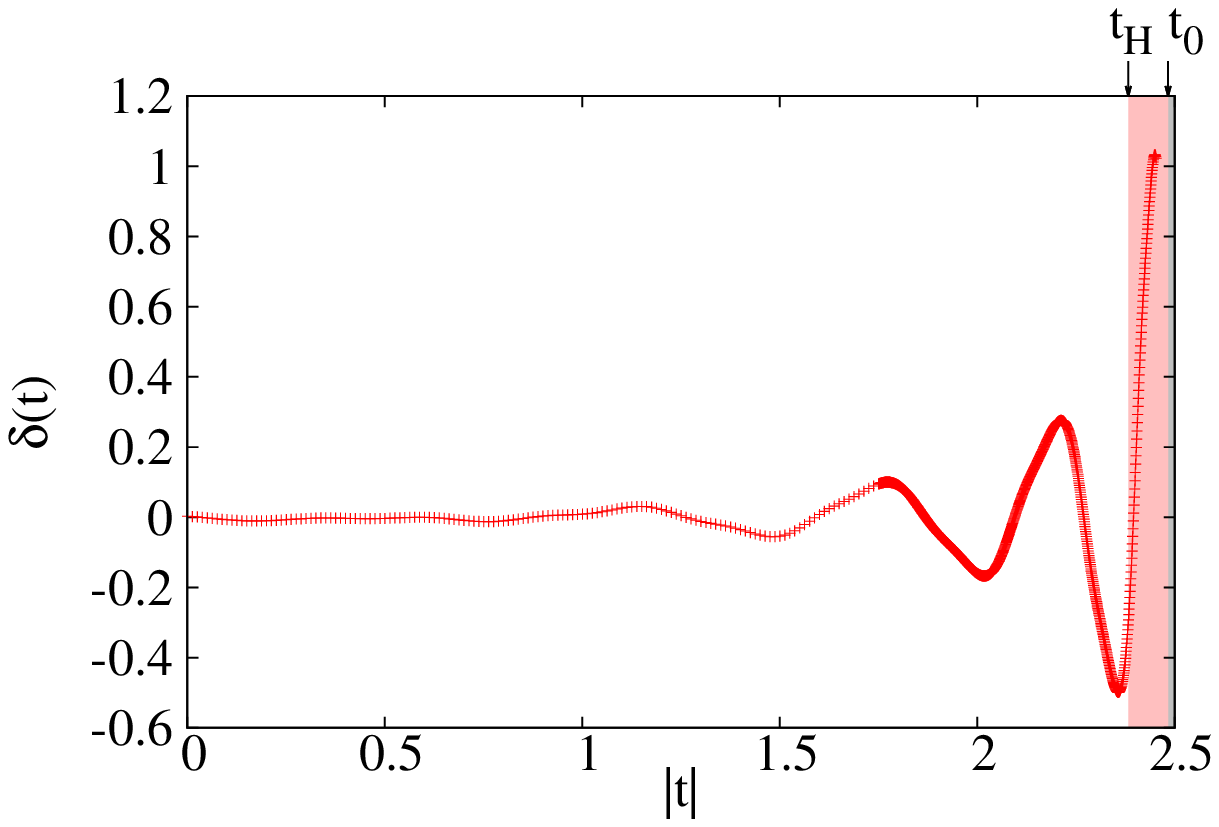}
\end{tabular}
\centering
{\footnotesize $\mathcal{A}=0.09$:$\phi=1.05$.}
\caption{
Evolution of the Fourier component $\gamma_{xx}(t,\vec{k}_{0})/a(t)$ (the left row) and the deviation \eqref{def-deviation} $\delta(t)$ (the right row). 
The integral constant $\phi$ in $h_{xx}(t,\vec{k}_{0})$ is 
determined by fitting the numerical data in the region $0.0<t<1.0$. 
We define the horizon crossing time $t_{H}$ and the big bang singularity time $t_{0}$ as $L^{\rm Isa}_{\rm V}(t_{H})=H(t_{H})^{-1}$ and $L^{\rm Isa}_{\rm V}(t_{0})=0$, respectively. 
}\label{Fourier transformation}
\end{figure}

We introduce the deviation $\delta(t)$ as follows 
\begin{equation}\label{def-deviation}
\delta(t)\equiv2\frac{\gamma_{xx}(t,\vec{k}_{0})-a(t)^{2}h_{xx}(t,\vec{k}_{0})}{\mathcal{A}a(t)}. 
\end{equation}
The evolution of $\gamma_{xx}(t,\vec{k}_{0})/a(t)$, $a(t)h_{xx}(t,\vec{k}_{0})$ and $\delta(t)$ 
are shown in Fig.~\ref{Fourier transformation} for 
$\mathcal{A}=0.1$, $\mathcal{A}=0.11$ and $\mathcal{A}=0.09$. 
Around the horizon crossing, 
the behavior of $\delta(t)$ 
depends on 
the value of $\mathcal{A}$. 
So, one may expect that 
the deviation from Isaacson's formula depends on $\mathcal{A}$. 
Although 
the WKB approximation is 
violated around the horizon crossing, the values of $\delta_{\rm V}$ and $\delta_{\rm A}$ 
do not depend so much on the value of $\mathcal A$. 
On the other hand,  
we can see significant dependence of $\mathcal A$ on $\delta_{\rm L}$. 
In particular, on the line $2$ and $6$, which correspond to anti-node 
of the standing waves, this dependence is large 
(for example, for $\mathcal{A}=0.07, 0.08, 0.09, 0.1, 0.11$, ${\rm max}(|\delta_{\rm L2}(H \sim 1/L_{\rm V})|)$ is $60\%$, $40\%$, $60\%,$ $30\%$, and $40\%$, respectively). 
\subsection{Temporal average}\label{sec temporal average}
\begin{figure}
\centering{
\includegraphics[width=7cm]{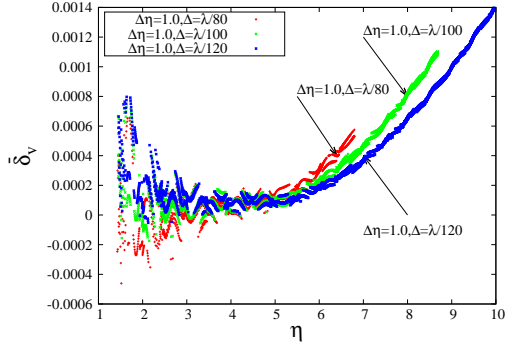}
}
\caption{
Evolution of $\bar{\delta}_{V}(\eta)=\frac{\langle L_{\rm V}\rangle_{\rm tem}(\eta)-L_{\rm V}^{\rm Is}(\eta)}{L_{\rm V}^{\rm Is}(\eta)}$ 
for different grid sizes.
The red, green and blue lines 
are 
$\Delta=\lambda_{0}/80$, $\Delta=\lambda_{0}/100$ and $\Delta=\lambda_{0}/120$, respectively. 
The parameter $b$ in the $L_{\rm V}^{\rm Is}=\sqrt{2b t+1}$ is 0.223, 
which is given by fitting the numerical data in the region $0.8<L_{\rm V}/\lambda_{0}<1$. 
}\label{graph temporal average}
\end{figure}

As is mentioned in 
Sec.~\ref{Sec Physical quantities}, to compare our results with 
the original Isaacson's formula, 
it is necessary to consider not only spatial average but also temporal one. 
The necessity of the temporal average is also 
explicitly shown in Appendix~\ref{Isaacson_in_Freidmann} for Isaacson's formula in the FLRW universe. 
However, because of the lack of the temporal periodicity, 
the temporal averaging is fairly ambiguous in the present case. 
We consider the following temporal averaging: 
\begin{equation}\label{temporal average}
\langle L_{\rm V}\rangle_{\rm tem}(\eta)
\equiv\frac{1}{\lambda_{0}}\int_{\eta-\lambda_{0}/2}^{\eta+\lambda_{0}/2}d\eta L_{\rm V},
\end{equation}
where $\eta=\int^{t}dt/a$ is the conformal time. 
We note that the comoving wavelength of the linear gravitational waves 
is $\lambda_{0}$. 
For simplicity, we use the scale factor of the radiation dominated universe 
$a=\sqrt{2b t+1}$
\footnote{
In this discussion, 
we always consider expanding universe, thus
the scale factor $a$ is an increasing function of time. 
}
as the scale factor in the definition of $\eta$, 
where $b$ is given by fitting $\sqrt{2b t+1}$ to 
$L_{\rm V}/\lambda_{0}$ in the region $L_{\rm V}/\lambda_{0}=[0.8,1.0]$. 
Thus, the conformal time is rewritten as the following form: 
$\eta=a/b=1/\sqrt{b H}$. 
The deviation 
$\bar{\delta}_{\rm V}=\frac{\langle L_{\rm V}\rangle_{\rm tem}(\eta)-L_{\rm V}^{\rm Is}(\eta)}{L_{\rm V}^{\rm Is}(\eta)}$ 
is depicted as a function of $\eta$ in 
Fig.~\ref{graph temporal average}. 
The value of $\bar{\delta}_{\rm V}$ does not converge within our 
numerical precision because the $\bar{\delta}_{\rm V}$ depends on $\Delta$. 
Nevertheless, from this figure, 
it is suggested that the deviation 
$\bar{\delta}_{\rm V}$ is at most $\sim0.1\%$. 
\section{Summary and discussion}
In this paper, we have investigated validity of Isaacson's formula 
by solving the Einstein equations with technique of numerical relativity. 
By solving the Hamiltonian constraint, we numerically constructed 
the initial data of the universe which contains the nonlinear 
standing gravitational waves in a cubic box with the periodic boundary. 
Then the time evolution was performed with the COSMOS code based on the BSSN formalism.
In order to investigate the validity of Isaacson's formula, we calculated 
the relation between the effective scale factor and the Hubble parameter 
and compared it with Isaacson's formula. 
The effective scale factors are defined 
from the proper volume, proper area and proper length, and the Hubble parameter 
is defined from trace of the extrinsic curvature of the time slice.
The results are summarized in Table~\ref{table of difference}. 
\begin{table}[htbp]
\begin{center}
\begin{tabular}{|c||c|c|c|}
\hline
&$L \sim 1/H$&$L \sim 0.5/H$&$L \sim 0.1/H$\\
\hline
\hline
volume&$\delta_{\rm Vmax}\sim3\%$&$\delta_{\rm Vmax}\sim2\%$&$\delta_{\rm Vmax}\sim0.5\%$\\
\hline
area&$\delta_{\rm Amax}\sim4\%$&$\delta_{\rm Amax}\sim2\%$&$\delta_{\rm Amax}\sim0.5\%$\\
\hline
length&$\delta_{\rm Lmax}\sim60\%$&$\delta_{\rm Lmax}\sim30\%$&$\delta_{\rm Lmax}\sim5\%$\\
\hline
\end{tabular}
\end{center}
\caption{
The largest value of the deviations from Isaacson's formula 
are listed for each value of 
the effective scale factors $L$ determined by the proper volume, area and length. 
} 
\label{table of difference}
\end{table}
The deviation from Isaacson's formula~$\delta_{\rm V}$ is at most about 3\% 
in the range of our numerical calculation given by 
$L\lesssim 1/H$. 
Although Isaacson's formula is not guaranteed 
for 
$L\sim 1/H$, 
this formula has still a few \% accuracy. 
Behavior of $L_{A0}(H)$, $L_{A1}(H)$ and $L_{A2}(H)$ are 
similar to that of the proper volume. 
While, for the effective scale factor defined by the proper length, deviation from the Isaacson's formula 
depends on the line position. 
This means that, when we use the proper length as the effective scale factor, 
the position dependence must be carefully treated. 
We also discussed the temporal average of the effective scale factor 
given by the proper volume, which gives a central value of 
the temporal oscillation of the effective scale factor. 
Our calculation of the temporal average suggests that 
the deviation of the central value from Isaccson's formula is at most 
$\sim 0.1\%$ in the range of our calculation. 
\par
One might expect that, when the gravitational wavelength is much longer than 
the Hubble scale, our model approaches to the Kasner solution. 
However, in our calculation, we could not proceed time evolution beyond 
$L \sim 1.5/H$. 
At this time, Hamiltonian constraint is violated, 
and we could not see any typical feature of the Kasner solution. 
Even using the finer resolution, the violation time does not change. 
This time $t_{c}$ of the constraint violation is near 
the big bang singularity time $t_{0}=\frac{1}{2\alpha}$ (see Eq.~(\ref{scale factor})): 
\begin{equation}
| t_{0}-t_{c} | \sim \frac{0.3}{H} \ \ \ \ (\mbox{in the case of } \mathcal{A}=0.1).
\end{equation}
Therefore, this constraint violation would be originated from the big bang singularity. 
We need more refined numerical method to analyze the structure of the spacetime around the singularity\cite{Garfinkle:2003bb}. 
Introducing another time coordinate, such as the e-folding number, 
we may investigate more details of the behavior in the early stage of the universe. 
We leave it as a future work. 


\section*{Acknowledgements}
We are grateful to Ken-ichi~Nakao and Misao~Sasaki for helpful discussions and comments. 

\section*{Appendix}

\appendix
\section{Time evolution equations}\label{part-of-time-evolution}
Here, we summarize time evolution part of the Einstein equation in vacuum. 
\begin{eqnarray}
\left( \frac{\partial}{\partial t}-\mathcal{L}_{\beta} \right) \Psi
&=&\frac{\Psi}{6}\left(\tilde{D}_{i}\beta^{i}-\alpha K\right), 
\\
\left( \frac{\partial}{\partial t}-\mathcal{L}_{\beta} \right) \tilde{\gamma}_{ij}&=&-2\alpha\tilde{A}_{ij}-\frac{2}{3}\tilde{D}_{k}\beta^{k}\tilde{\gamma}_{ij},
\\
\left(\frac{\partial}{\partial t}-\mathcal{L}_{\beta}\right)K&=&-\Psi^{-4}(\tilde{D}_{i}\tilde{D}^{i}\alpha+2\tilde{D}_{i}\ln \Psi \tilde{D}^{i}\alpha)+\alpha[\tilde{A}_{ij}\tilde{A}^{ij}+\frac{K^{2}}{3}],~
\\
\left(\frac{\partial}{\partial t}-\mathcal{L}_{\beta}\right)\tilde{A}_{ij}&=&
-\frac{2}{3}\tilde{D}_{k}\beta^{k}\tilde{A}_{ij}+\alpha\left( K\tilde{A}_{ij}-2\tilde{\gamma}^{kl}\tilde{A}_{ik}\tilde{A}_{jl} \right) \nonumber
\\
&&+\Psi^{-4}\{ -\tilde{D}_{i}\tilde{D}_{j}\alpha+2\tilde{D}_{i} \ln \Psi \tilde{D}_{j}\alpha+2\tilde{D}_{j}\ln \Psi \tilde{D}_{i}\alpha \nonumber
\\
&&+\frac{1}{3}(\tilde{D}_{k}\tilde{D}^{k}\alpha-4\tilde{D}_{k} \ln \Psi \tilde{D}^{k} \alpha)\tilde{\gamma}_{ij} \nonumber
\\
&&+\alpha [ \tilde{R}_{ij}-\frac{1}{3}\tilde{R}\tilde{\gamma}_{ij}-2\tilde{D}_{i}\tilde{D}_{j}\ln \Psi+4\tilde{D}_{i} \ln \Psi \tilde{D}_{j} \ln \Psi \nonumber
\\
&&+\frac{2}{3}(\tilde{D}_{k}\tilde{D}^{k} \ln \Psi-2\tilde{D}_{k}\ln \Psi \tilde{D}^{k}\ln \Psi)\tilde{\gamma}_{ij} ]\}, 
\end{eqnarray}
where $\alpha$ is the lapse function, 
$\beta$ is the shift vector, $\mathcal{L}_{\beta}$ is the Lie derivative with respect to 
$\beta$, and the other variable defined in Sec.\ref{Set up of initial data}. 
\section{Isaacson's formula in Friedmann universe}\label{Isaacson_in_Freidmann}
Isaacson investigated the effective energy momentum tensor of 
low amplitude and high frequency gravitational 
waves\cite{Isaacson:1967zz,Isaacson:1968zza} in general background. 
In this appendix, we derive the effective energy-momentum tensor of 
low amplitude and high frequency gravitational waves 
in the spatially flat FLRW universe. 
The full metric can be written as 
\begin{equation}
ds^{2}=-dt^{2}+a^{2}(t)\left\{\delta_{ij}+h_{ij}(t,\vec{x})\right\} dx^{i}dx^{j}, 
\end{equation}
where $a(t)$ is the scale factor and $h_{ij}$ is low amplitude and high frequency perturbation which satisfies the transverse and traceless gauge conditions: 
$\delta^{ij}h_{ij}=0$, $\partial^{i}h_{ij}=0$. 
\par
In order to consider the low amplitude and high frequency gravitational waves, 
we introduce a small parameter $\epsilon\ll1$ and 
assume the following order of perturbation: 
$h_{ij}\sim \mathcal{A} \sim \lambda/l \sim \epsilon$, 
where $\mathcal{A}$ is typical amplitude of gravitational waves $\lambda$ is typical wavelength of gravitational waves, 
$l$ is a typical scales of background Hubble scale in the present case. 
Then, we find $\partial_k h_{ij}\sim \epsilon^0$ and 
$\partial_{l}\partial_{k} h_{ij} \sim \epsilon^{-1}$. 
\par
We decompose the Ricci tensor by the order of $\epsilon$ as follows: 
\begin{equation}
R_{\mu\nu}=\bar{R}^{(0)}_{\mu\nu}+R^{(-1)}_{\mu\nu}+R^{(0)}_{\mu\nu}+\mathcal O(\epsilon), 
\end{equation}
where we assign $\bar{R}^{(0)}_{\mu\nu}=\mathcal{O}(\epsilon^{0})$ for the background part. 
Each terms are 
\begin{eqnarray}
\bar{R}^{(0)}_{00}&=&-\frac{3\ddot{a}}{a},\\
\bar{R}^{(0)}_{0i}&=&0,\\
\bar{R}^{(0)}_{ij}&=&(\ddot{a}a+2\dot{a}^{2})\delta_{ij},\\
R^{(-1)}_{00}&=&0,\\
R^{(-1)}_{0i}&=&0,\\
R^{(-1)}_{ij}&=&
\frac{a^{2}}{2}\left\{ \ddot{h}_{ij}
-\frac{1}{a^{2}}\nabla^{2}h_{ij}
+3H\dot h_{ij}\right\},\\
R^{(0)}_{00}&=&
\frac{1}{4a^{2}}\dot{h}_{ij}\dot{h}^{ij}
+\frac{1}{2a^{2}}\ddot{h}_{ij}h^{ij},
\label{R200}\\
R^{(0)}_{0i}&=&
\frac{1}{4a^{2}}\dot{h}_{kl}\partial_{i}h^{kl}
+\frac{1}{2a^{2}}h_{kl}\partial_{i}\dot{h}^{kl}
-\frac{1}{2a^{2}}h^{mn}\partial_{m}\dot{h}_{ni} , \\
R^{(0)}_{ij}&=&\frac{1}{4a^{2}}\partial_{j}h^{mn}\partial_{i}h_{mn}
+\frac{1}{2a^{2}}h^{mn}\partial_{i}\partial_{j}h_{mn}
+\frac{1}{2a^{2}}h^{mn}\partial_{m}\partial_{n}h_{ij}
\cr
&-&\frac{1}{2a^{2}}h^{mn}\partial_{m}\partial_{j}h_{ni}
-\frac{1}{2a^{2}}h^{mn}\partial_{m}\partial_{i}h_{nj}
-\frac{1}{2}\dot{h}^{m}_{~j}\dot{h}_{mi}
\cr
&+&\frac{1}{2a^{2}}\partial^{n}h^{m}_{~j}\partial_{n}h_{mi}
-\frac{1}{2a^{2}}\partial^{n}h^{m}_{~j}\partial_{m}h_{ni},
\label{R2ij}
\end{eqnarray}
and dot denotes the time derivative. 
\par
In our ordering of the perturbation, 
the leading order equation is 
given by 
\begin{equation}\label{Einstein-eq-wave}
\frac{\partial^2}{\partial t^2} h_{ij}(t,\vec{x})+3H\frac{\partial}{\partial t} h_{ij}(t,\vec{x})-\frac{1}{a^2}\nabla^2h_{ij}(t,\vec{x})=0. 
\end{equation} 
Using the WKB approximation, the solution is 
\begin{equation}
h_{ij}(t,\vec{x})=\frac{1}{a}\int\frac{l^{3}d^{3}k}{(2\pi)^{3}}(\mathcal{A}^{(\vec{k})}_{ij}u_{\vec{k}}(t)e^{i\vec{k}\cdot\vec{x}}+\mbox{c.c.}),
\label{solpsi}
\end{equation}
where $u_{\vec{k}}(t)$ is defined by
\begin{equation}
u_{\vec{k}}(t)=e^{i\int^{t}_{0}\frac{k}{a}dt},
\end{equation}
$l$ is the size of the comoving box 
and $\mathcal{A}_{ij}^{(\vec{k})}$ is an integration constant which satisfies $\delta^{ij}\mathcal{A}_{ij}^{(\vec{k})}=0$ and $k^{i}\mathcal{A}^{(\vec{k})}_{ij}=0$. 
\par
In the next leading terms of the Einstein equations, 
we find the time evolution equation of the background. 
Since we are interested in the backreaction effect of $h_{\mu\nu}$ 
on the dynamics of the background metric $\bar{g}_{\mu\nu}$, 
we extract homogeneous and isotropic part of contribution from $R_{\mu\nu}$. 
That is, we consider the following field equation 
\begin{equation}
\bar{R}^{(0)}_{\mu\nu}-\frac{1}{2}\bar{R}^{(0)}\bar{g}_{\mu\nu}=8\pi T^{({\rm GW})}_{\mu\nu},
\end{equation}
with the effective energy-momentum tensor is defined by 
\begin{equation}
T^{({\rm GW})}_{\mu\nu}=
-\frac{1}{8\pi}(\langle R^{(0)}_{\mu\nu}\rangle -\frac{1}{2}\bar{g}_{\mu\nu}\langle R^{(0)}\rangle), 
\end{equation}
where $\langle \rangle$ denotes a spatial average which smoothes out the inhomogeneities and 
extracts the spatially homogeneous and isotropic part.
The average scale is assumed to be larger than the wavelength of gravitational waves 
and we explicitly describe how to take this average below. 
\par
First, 
to preserve the covariance of each variables  
under the spatial rotation, 
we impose the following property for the average for any physical quantity $Q(\vec k,\vec k')$:
\begin{eqnarray}
\langle Q(\vec k, \vec k') u_{\vec{k}}u_{\vec{k'}}(t,x)\rangle
&=&\langle Q(\vec k, \vec k')e^{i\int^{t}_{0}\frac{k+k'}{a}dt}e^{i(k_{j}+k'_{j})\cdot x_{j}}\rangle\cr
&=&\langle Q(\vec k, -\vec k)\rangle \frac{(2\pi)^{3}}{l^{3}}\delta(\vec{k}+\vec{k'})e^{i2\int^{t}_{0}\frac{k}{a}dt},
\end{eqnarray}
and 
\begin{eqnarray}
\langle Q(\vec k, \vec k') u_{\vec{k}}^{\ast}u_{\vec{k'}}(t,\vec{x})\rangle
&=&\langle Q(\vec k, \vec k') e^{i\int^{t}_{0}\frac{-k+k'}{a}dt}e^{i(-k_{j}+k'_{j})\cdot x_{j}}\rangle\cr
&=&\langle Q(\vec k, \vec k) \rangle \frac{(2\pi)^{3}}{l^{3}}\delta(\vec{k}-\vec{k'}).
\end{eqnarray}
Substituting Eqs.~\eqref{solpsi} into Eq.~\eqref{R200} and \eqref{R2ij}, 
we obtain 
\begin{eqnarray}
\langle R^{(0)}_{00}\rangle&=&-\int\frac{l^{3}d^{3}k}{(2\pi)^{3}}
\Bigl\{\frac{3k^2}{4}\left(\mathcal A^{(\vec k)}_{ij}\mathcal A^{(-\vec k)}_{ij}
e^{2i\int^{t}_{0}\frac{k}{a}t}+\mathcal A^{(\vec k)*}_{ij}\mathcal A^{(-\vec k)*}_{ij}
e^{-2i\int^{t}_{0}\frac{k}{a}dt}\right)\cr
&&\hspace{2cm}
+\frac{k^2}{2a^4}\left|\mathcal A^{(-\vec k)}_{ij}\right|^2\Bigr\}, \\
\langle R^{(0)}_{ij}\rangle&=&-\langle
\int\frac{l^{3}d^{3}k}{(2\pi)^{3}}
\Bigl\{
\frac{k_ik_j}{4a^2}\left(\mathcal A^{(\vec k)}_{kl}\mathcal A^{(-\vec k)}_{kl}
e^{2i\int^{t}_{0}\frac{k}{a}dt}
+\mathcal A^{(\vec k)*}_{kl}\mathcal A^{(-\vec k)*}_{kl}
e^{-2i\int^{t}_{0}\frac{k}{a}dt}\right)
\cr
&&-\frac{k^2}{a^2}\left(\mathcal A^{(\vec k)}_{ik}\mathcal A^{(-\vec k)}_{jk}
e^{2i\int^{t}_{0}\frac{k}{a}dt}
+\mathcal A^{(\vec k)*}_{ik}\mathcal A^{(-\vec k)*}_{jk}
e^{-2i\int^{t}_{0}\frac{k}{a}dt}\right)
\cr
&&-\frac{3k_ik_j}{2a^2}
\left|\mathcal A^{(\vec k)}_{kl}\right|^2
+\frac{k^2}{a^{2}}\left(\mathcal A^{(\vec k)}_{ik}\mathcal A^{(\vec k)*}_{jk}
+\mathcal A^{(\vec k)*}_{ik}\mathcal A^{(\vec k)}_{jk}\right)\Bigr\}
\rangle. 
\end{eqnarray}
The Ricci scalar $R^{(0)}$ is given by 
\begin{equation}
\langle R^{(0)}\rangle =\frac{3}{2a^4}
\int\frac{l^{3}d^{3}k}{(2\pi)^{3}}k^2\left(
\mathcal A^{(\vec k)}_{ij}\mathcal A^{(-\vec k)}_{ij}e^{2i\int^{t}_{0}\frac{k}{a}dt}
+\mathcal A^{(\vec k)*}_{ij}\mathcal A^{(-\vec k)*}_{ij}e^{-2i\int^{t}_{0}\frac{k}{a}dt}\right). 
\end{equation}
Then, $T_{00}^{({\rm GW})}$ is given by 
\begin{equation}\label{T00}
T_{00}^{({\rm GW})}=\frac{1}{2a^4}
\int\frac{l^{3}d^{3}k}{(2\pi)^{3}}k^2
\left|\mathcal A^{(\vec k)}_{ij}\right|^2. 
\end{equation} 
By extracting the trace part, we obtain 
\begin{eqnarray}\label{Tij2}
T_{ij}^{({\rm GW})}&=&
\frac{\delta_{ij}}{2a^2}
\int\frac{l^{3}d^{3}k}{(2\pi)^{3}}k^2
\Bigl\{
\left(
\mathcal A_{kl}^{(\vec k)}\mathcal A_{kl}^{(-\vec k)}
e^{2i\int^{t}_{0}\frac{k}{a}dt}
+
\mathcal A_{kl}^{(\vec k)*}\mathcal A_{kl}^{(-\vec k)*}
e^{-2i\int^{t}_{0}\frac{k}{a}dt}\right)\cr
&&\hspace{3cm}+\frac{1}{3}\left|\mathcal A^{(\vec k)}_{kl}\right|^2\Bigr\}. 
\end{eqnarray}
The first and the second terms inside the integration of Eq.~\eqref{Tij2} 
are temporally oscillating, and we may eliminate this term by 
taking a temporal average and get following form:
\begin{equation}\label{Tij3}
T_{ij}^{({\rm GW})}=
\frac{\delta_{ij}}{6a^2}
\int\frac{l^{3}d^{3}k}{(2\pi)^{3}}k^2
\left|\mathcal A^{(\vec k)}_{kl}\right|^2. 
\end{equation}
According to \eqref{T00} and \eqref{Tij3} , we obtain traceless energy momentum tensor like 
the radiation fluid. 
It means that the energy density of the effective energy momentum tensor 
is proportional to $1/a^{4}$. 



\begin{thebibliography}{99}
  \bibitem{Ellis:2011hk} 
  G.~F.~R.~Ellis,
  {\it Class.\ Quantum Grav.}\  {\bf 28}, 164001 (2011)
  
   \bibitem{Green:2014aga} 
  S.~R.~Green and R.~M.~Wald,
  {\it Class.\ Quantum\ Grav.}\  {\bf 31}, 234003 (2014)
  
\bibitem{Nambu:2000ex}
  Y.~Nambu,
  {\it Phys.\ Rev.\ D} {\bf 62} (2000) 104010
  
\bibitem{Kai:2006ws} 
  T.~Kai, H.~Kozaki, K.~i.~Nakao, Y.~Nambu and C.~M.~Yoo,
  {\it Prog.\ Theor.\ Phys.}\  {\bf 117}, 229 (2007)
  
  \bibitem{Isaacson:1967zz} 
  R.~A.~Isaacson,
  {\it Phys.\ Rev.}\  {\bf 166}, 1263 (1967).
  
  \bibitem{Isaacson:1968zza} 
  R.~A.~Isaacson,
  {\it Phys.\ Rev.}\  {\bf 166}, 1272 (1968).
  
   \bibitem{Bentivegna:2012ei} 
  E.~Bentivegna and M.~Korzynski,
  {\it Class.\ Quantum\ Grav.}\  {\bf 29}, 165007 (2012)
  
  \bibitem{Yoo:2012jz} 
  C.~M.~Yoo, H.~Abe, K.~i.~Nakao and Y.~Takamori,
  {\it Phys.\ Rev.\ D} {\bf 86}, 044027 (2012)
  
\bibitem{Bentivegna:2013jta} 
  E.~Bentivegna and M.~Korzynski,
  {\it Class.\ Quantum\ Grav.}\  {\bf 30}, 235008 (2013)
   
  \bibitem{Yoo:2014boa} 
  C.~M.~Yoo and H.~Okawa,
  {\it Phys.\ Rev.\ D} {\bf 89}, no. 12, 123502 (2014)
  
    \bibitem{Nakao}
  K.~i.~Nakao private communication,
  
    \bibitem{Shibata:1995we} 
  M.~Shibata and T.~Nakamura,
  {\it Phys.\ Rev.\ D} {\bf 52}, 5428 (1995).
  
    \bibitem{Baumgarte:1998te} 
  T.~W.~Baumgarte and S.~L.~Shapiro,
  {\it Phys.\ Rev.\ D} {\bf 59}, 024007 (1999)
  
\bibitem{Yamamoto:2008js} 
  T.~Yamamoto, M.~Shibata and K.~Taniguchi,
  {\it Phys.\ Rev.\ D {\bf 78}, 064054 (2008)}
  
  \bibitem{Yoo:2013yea} 
  C.~M.~Yoo, H.~Okawa and K.~i.~Nakao,
  {\it Phys.\ Rev.\ Lett.}\  {\bf 111}, 161102 (2013)
  
    \bibitem{Clifton:2009jw} 
  T.~Clifton and P.~G.~Ferreira,
  {\it Phys.\ Rev.\ D} {\bf 80}, 103503 (2009)
  [{\it Erratum-ibid.\ D} {\bf 84}, 109902 (2011)]
  
  \bibitem{Clifton:2012qh} 
  T.~Clifton, K.~Rosquist and R.~Tavakol,
  {\it Phys.\ Rev.\ D} {\bf 86}, 043506 (2012)
  
\bibitem{Garfinkle:2003bb} 
  D.~Garfinkle,
  {\it Phys.\ Rev.\ Lett.}\  {\bf 93}, 161101 (2004)
  
  
  \end{thebibliography}
\end{document}